\documentclass[reqno,11pt,twoside]{article} 
\usepackage{color}

\usepackage{amsmath} 
\usepackage{amssymb}
\usepackage{latexsym}
\usepackage{cite}

\newcommand{\Mark}[1]{({\em #1}) \marginpar{{\bf !}}}

\def\undertilde#1{\mathord{\vtop{\ialign{##\crcr
$\hfil\displaystyle{#1}\hfil$\crcr\noalign{\kern1.5pt\nointerlineskip}
$\hfil\tilde{}\hfil$\crcr\noalign{\kern1.5pt}}}}}
\def\wideundertilde#1{\mathord{\vtop{\ialign{##\crcr
$\hfil\displaystyle{#1}\hfil$\crcr\noalign{\kern1.5pt\nointerlineskip}
$\hfil\widetilde{}\hfil$\crcr\noalign{\kern1.5pt}}}}}

\newcommand{\RR}{\mathbb{R}}

\newcommand{\NN}{\mathbb{N}}
\newcommand{\CC}{\mathbb{C}}

\newenvironment{Proof}%
{\par \medskip \noindent {\em Proof.}}{\hspace*{\fill} $\square$\par%
\medskip\noindent}
{\par \medskip \noindent {\em Proof\, }}{\hfill $\square$ 
\medskip }
\newtheorem{Thm}{Theorem}
\newtheorem{Prop}[Thm]{Proposition}
\newtheorem{Lem}[Thm]{Lemma}

\newtheorem{Def}[Thm]{Definition}

\newcommand{\bs}[1]{\boldsymbol{#1}}

\renewcommand{\Im}{\text{Im }}
\newcommand{\lsp}{\left(\,}     
\newcommand{\rsp}{\,\right)}

\newcommand{\calA}{\mathcal{A}}          
\newcommand{\calT}{\mathcal{T}}          
\newcommand{\calH}{\mathcal{H}}
\newcommand{\calD}{{\mathcal D}}
\newcommand{\calO}{{\mathcal O}}
\newcommand{\calS}{{\mathcal S}}

\newcommand{\half}{{\frac{1}{2}}}

\renewcommand{\d}{{\rm d}}

\newcommand{\eps}{\varepsilon}
\newcommand{\bfp}{{\boldsymbol{p}}}
\newcommand{\unity}{{\setlength{\unitlength}{1em}
                     \begin{picture}(0.75,1)
                     \put(0,0){$1$}
                     \put(0.34,0){\line(0,1){0.65}}
                     \end{picture}
                   }}
\newcommand{\einsi}{\frac{1}{i}}

\newcommand{\unit}{{\mbox{\texttt 1}}} 

\newcommand{\Spd}{H}                
\newcommand{\spd}{e}                   
\newcommand{\Spdc}{\Spd^{\rm c}}  
\newcommand{\Tub}{{\mathcal T}_+}  
\newcommand{\spdc}{\spd}  
\newcommand{\spdr}{\spd'} 
\newcommand{\spdi}{\spd''} 

\newcommand{\Po}{P_+^{\uparrow}}

\newcommand{\Lor}{L_+^{\uparrow}}
\newcommand{\LorOrtho}{L^{\uparrow}}

\renewcommand{\lor}{\Lambda}     
\newcommand{\Boo}[2]{\lor_{#1}(#2)}         
\newcommand{\WigRot}{R}       

\newcommand{\pt}{\text{\rm p}}
\newcommand{\st}{\text{\rm s}}
\newcommand{\krein}{\text{\rm K}}
\newcommand{\IntFctPt}{v^\pt}    
\newcommand{\IntFct}{v}         
\newcommand{\IntFctC}{v^c}  
\newcommand{\IntFctPtC}{{v^{\pt c}}}   
\newcommand{\IntFctAPt}{v^{\pt}}
\newcommand{\IntFctA}{v} 
\newcommand{\IntFctF}{v^{F}}      
\newcommand{\IntFctPhi}{v^{\prime}} 

\newcommand{\IntFctsPt}{v^{s,\pt}}    
\newcommand{\IntFcts}{v^{s}}         
\newcommand{\IntFctsF}{v^{s,F}}      
\newcommand{\IntFctsEsc}[1]{v^{(s,#1)}}  

\newcommand{\MTwoPtpt}{\MTwoPt^\pt}
\newcommand{\MTwoPt}{M}    
  
\newcommand{\field}{\varphi}
\newcommand{\String}{S}    
\newcommand{\Hyp}{H_m^+}
\newcommand{\Uirr}{U^{(m,s)}} 
\newcommand{\Ds}{D^{(s)}}

\newcommand{\Hirr}{\calH^{(m,s)}}
\newcommand{\Hlittle}{\mathfrak{h}}                
\newcommand{\HlittleS}{\Hlittle^{(s)}}                
     
\newcommand{\h}{h}  
\newcommand{\hpt}{h^\pt}  
\newcommand{\D}{D}      

\newcommand{\Eps}{E_+}
\newcommand{\sz}{k}
\newcommand{\PT}{{\rm PT}}    
\newcommand{\Done}{D^{(1)}}
\newcommand{\target}{\mathfrak{h}}     
\newcommand{\Uone}{U^{(m,1)}} 
\newcommand{\APt}{A^\pt} 
\newcommand{\ASt}{A}        
\newcommand{\AK}{A^\krein}        
\newcommand{\phiK}{\phi^\krein}        
\newcommand{\uK}{u^\krein}        
\newcommand{\tildeuK}{\tilde{u}^\krein}        
\newcommand{\de}{d_e}        
\newcommand{\PotPt}{{A^\pt}} 
\newcommand{\F}{F}          
\newcommand{\Pot}{A}        

\newcommand{\Lint}{L}
\newcommand{\LintPt}{{L^\pt}}
\newcommand{\LintSt}{{L^\st}}
\newcommand{\LintK}{{L^\krein}}
\newcommand{\Dirac}{\psi}
\newcommand{\Wickli}{:\!}
\newcommand{\Wickre}{\!:}
\newcommand{\TwoPt}{w}
\begin{document} 
\markboth{\scriptsize{J.\ Mund, E.\ T.\ de Oliveira}}{\scriptsize{
    String--Localized Massive Bosons }} 
\title{String--localized free vector and tensor potentials for \\
massive particles with any spin: I. Bosons} 
\author{Jens Mund\footnote{e-mail: mund@ufjf.fisica.br} ${}^{1}$,
  Erichardson T.\ de Oliveira\footnote{e-mail: erifisico@yahoo.com.br}
  ${ }^{1,2}$}  
\date{${}^{1}$Dept.\ de F\'{\i}sica, Universidade Federal de Juiz de
  Fora, Brazil\\
${}^{2}$CEFET/RJ, Campus Valen\c ca, Brazil. \\[2ex] 
September 28, 2016
} 
\maketitle
\begin{abstract}
It is well-known that a (point-localized) free quantum field for
massive particles with spin $s$  
acting in a Hilbert space has at best scaling dimension $s+1$, which
excludes its use in the perturbative construction of 
renormalizable interacting models 
for higher spin ($s\geq 1$).   
Up to date, such models have been constructed only in the context of
gauge theory, at the cost of introducing additional unphysical (ghost)
fields and an unphysical (indefinite metric) state space. The
unphysical degrees of freedom are divided out by requiring gauge (or
BRST) invariance.   

We construct free quantum fields for higher spin particles which have the
same good UV behaviour as the scalar field (scaling dimension one),
and at the same time act on a Hilbert space without ghosts.  
They are localized on semi-infinite strings extending to space-like infinity, 
but are linearly related to their point-local counterparts. 
We argue that this is sufficient locality for a perturbative
construction of interacting models of the gauge theory type, 
with a string-independent S-matrix and point-localized
interacting observable fields. 
The usual principle of gauge-invariance is here replaced by the (deeper)
principle of locality.   
\end{abstract}
\tableofcontents
\section{Introduction}  
As is well-known, the combined requirements of locality (or Einstein
causality), positivity of states, and positivity of
the energy lead to the bad short distance behaviour of quantum fields
(UV singularities).  
This behaviour is quantified by the so-called scaling dimension of the
quantum field. 
The scaling dimension gets worse with increasing spin: 
In particular, a free (point-) local quantum field for massive
particles with spin $s$ acting in a Hilbert space 
has at best scaling dimension $s+1$. 
Among the infinity of free local quantum fields for massive particles
with spin $s$~\cite{Weinberg}, this optimal UV behaviour is achieved by 
the so-called tensor potential $\PotPt_{\mu_1\cdots \mu_s}$. This is a
symmetric tensor field of rank $s$, which is trace- and divergence
free, and is uniquely characterized by these
properties~\cite{Weinberg64}.   
The high scaling dimension ($s+1$) of this field excludes its use in the
perturbative construction of renormalizable interacting models for $s\geq 1$. 

For spin one and two, there do exist fields in the context of gauge
theory with the same good UV behaviour as the scalar spin-zero field,
namely scaling dimension one~\cite{ScharfGauge}. 
However, they need the introduction of an unphysical state space
(a Krein space with indefinite metric), as well as unphysical fields (ghosts).  
In the construction of interacting models, the unphysical degrees of
freedom have to be divided out in the end by requiring gauge (or
BRST) invariance of observables and of the S-matrix.  
Although this approach is quite successful -- it is the basis of the standard
model of elementary particles -- it has some unsatisfying features.  
Firstly, the introduction of unphysical degrees of freedom is against
the spirit of Ockham's razor. 
More severely, the 
construction of charged physical states or fields in interacting
models is quite involved. In particular, the interacting Dirac field,
minimally coupled to the (Krein space) vector boson and constructed in
a direct perturbative manner,  is unphysical in the sense that its two-point  
function does not have a probability interpretation since it is not
positive definite. 
To the best of our knowledge, this problem has not
been overcome in the massive case.\footnote{\label{ChargedFields}%
For massless vector bosons (photons), it has indeed been claimed to be
impossible to construct charged physical states in Krein space using
the standard metric~\cite{Zwanziger76} or even as limits of local
states in any weak topology~\cite{Steinmann}.  
However, two constructions of charged physical states have been proposed:     
 (a)  Morchio and Strocchi constructed such fields
in~\cite{Strocchi03} in a Krein space setting. The construction is
however quite subtle due to the weak topology. 
 (b) O. Steinmann constructed such fields in~\cite{Steinmann84}, using
 his own perturbative scheme, which is, however, not widely known. 
} 

In this article we construct, for every integer $s$, a
free tensor quantum field $\ASt_{\mu_1\ldots \mu_s}$ for massive
particles with spin $s$, which shares the same good UV behaviour
(scaling dimension one), and in 
addition acts in a Hilbert space without indefinite metric and without ghosts.  
The price we have to pay is that our fields are not point-local, but instead 
are localized, in a sense explained below, on Mandelstam strings
extending to space-like infinity~\cite{MSY,MSY2} (or light-like infinity, see
footnote~\ref{lightlike}).   
We then have to take care that there are (point-) local observables 
in the Borchers class of our string-localized fields, and that we have
sufficient locality for a perturbative construction of interacting models. To
this end, it is gratifying that our fields coincide with the
point-local fields $\APt_{\mu_1\ldots \mu_s}$ up to derivative terms
(which are responsible for the bad UV behaviour);  
therefore suitably chosen interaction Lagrangeans differ from
their point-local counterparts by a divergence, which according to
standard folklore should be irrelevant to physics.
(In fact, one of the authors has proved this in the example of massive
QED at lower orders~\cite{Mund_MassiveQED}.)  
Thus, our fields have both a good UV behaviour and sufficient locality such that
the perturbative construction of interacting models with predictive
power seems feasible. Moreover, we comply with Ockham's razor
in that we do not have any unphysical states or degrees of freedom. 
A more important and concrete  advantage of our approach (over the
Krein approach) is that it allows for a straightforward perturbative
construction of  physical charged fields. 
An interesting feature of our interacting Dirac field in massive QED
is that in the limit when the photon mass goes to zero it seems to
describe the electron as an infra-particle. These claims are
explained in the Outlook.  

It requires further investigation if the class of renormalizable models in
our approach differs from that of the BRST approach, and if the 
renormalizable models are equivalent.    
The question of renormalizability of string-localized models is under
investigation~\cite{CardosoMund}. 

Before presenting our results in more detail, we need to explain 
our notion of string-localized quantum fields. 
This concept has been introduced by Steinmann~\cite{St}, inspired by ideas of
Mandelstam~\cite{Mandelstam} and Dirac~\cite{Dirac55}. It has been
further refined by one of the authors together with Schroer and 
Yngvason in the articles~\cite{MSY,MSY2}, where
basically\footnote{In~\cite{MSY,MSY2} only fields with scalar
  transformation behaviour have been considered.
} 
the present notion has been coined: 
A {\em string-localized quantum tensor field} is a 
multiplett 
of operator-valued
distributions $\field_{\mu_1 \cdots \mu_k}(x,e)$, where $x$ is a point
in Minkowski space and $e$ is in the manifold of space-like directions
\begin{equation} \label{eqSpd}
\Spd :=\{ e\in\RR^4:\, e\cdot e=-1\}.
\end{equation}  
The space-like\footnote{\label{lightlike}The choice of 
{\em space}-like strings is motivated by the known fact 
that in every massive model charge-carrying field operators are
localizable in space-like cones~\cite{BuF}. It  seems, however, that 
our constructions go through also for {\em light}-like strings,
replacing $\Spd$ by the light cone. 
} 
 string (or ray) emanating from $x$ in the direction
$e$, $\String_{x,e} \doteq  x+\RR_0^+e$,  
is the localization region of $\field_{\mu_1 \cdots \mu_k}(x,e)$ in the sense of
compatibility of quantum observables: If the strings $\String_{x,e}$ and
$\String_{x',e''}$ are space-like separated for all $e''$ in an open
neighborhood of $e'$, then 
\begin{equation} \label{eqFieldLoc} 
[\field_{\mu_1\cdots\mu_k}(x,e),\field_{\mu_1'\cdots\mu_k'}(x',e')]=0.
\end{equation}
It is further required that the multiplett transform as a tensor under a
unitary representation $U$ of the proper orthochronous Poincar\'e group $\Po$: 
\begin{equation}  \label{eqCovTens} 
U(a,\Lambda)\, \field_{\mu_1\cdots\mu_k}(x,e) \, U(a,\Lambda)^{-1} =
\field_{\alpha_1\cdots\alpha_k}(a+\Lambda x,
\Lambda e) \; \Lambda^{\alpha_1}{}_{\mu_1} \cdots
\Lambda^{\alpha_k}{}_{\mu_k}, 
\end{equation}
where $a\in\RR^4$ is a translation and $\lor$ is a Lorentz
transformation. (We use Einstein's sum convention in repeated Lorentz
indices.)  
We say that the field is a free field for a given particle type if it
creates from the vacuum only single particle states of the given
type. (A particle type corresponds to a unitary irreducible
representation of the 
Poincar\'e group, characterized by the values for mass and spin.)     
We consider here only massive bosons, {\em i.e.\ }particles with
non-zero mass, $m>0$, and integer spin, $s\in\NN_0$. 
By the {\em scaling dimension} of a string-localized quantum field we
mean one half of the scaling degree of its two-point function with
respect to coinciding $x$-arguments after smearing in the
$e$-variables, see section~\ref{secScalDeg} for details.   
We can now summarize our findings as follows. 
\begin{Thm}  \label{MainThm}
For every massive boson $(m,s)$ with $s\geq 1$, there is a free
string-localized quantum tensor field 
$\Pot_{\mu_1\cdots\mu_s}(x,e)$ acting in the Hilbert space of its
point-local counterpart $\PotPt_{\mu_1\cdots\mu_s}(x)$,
with the following properties:  

$(i)$ It has scaling dimension one 
  after smearing in the variable $e$; 

$(ii)$ It differs from its point-local counterpart $\PotPt_{\mu_1\cdots\mu_s}$ by
derivative terms. 
\\
More specifically, there are string-localized tensor fields
$\phi^{(s,0)},\ldots, \phi^{(s,s-1)}$, where $\phi^{(s,k)}$ has rank $k$,
such that 
\begin{align} \label{eqPotPotPts}
\Pot_{\mu_1\cdots \mu_s}(x,e)&=\PotPt_{\mu_1\cdots \mu_s}(x)+ 
\sum_{\substack{I\subset\{1,\ldots,s\}\\I\neq\emptyset}} 
\partial_{\mu_{i_1}}\cdots \partial_{\mu_{i_{|I|}}}
\phi^{(s,|I^c|)}_{\mu_{j_1}\cdots \mu_{j_{|I^c|}}}(x,e), 
\end{align}
where we have written $I=\{i_1,\ldots,i_{|I|}\}$ and 
$I^c=\{j_1,\ldots,j_{|I^c|}\}=$ complement of $I$. 
These are free fields for the same particle type $(m,s)$, acting in
the same Hilbert space. 
 
$(iii)$ All fields $\PotPt_{\mu_1\cdots \mu_s}$, $\Pot_{\mu_1\cdots \mu_s}$,
$\phi^{(s,k)}_{\mu_1\cdots\mu_k}$, are 
string-local relative to each other.  
Further, the fields $\Pot_{\mu_1\cdots \mu_s}$ have a massless
limit.\footnote{That is to say, their two-point functions have massless
  limits. Recall that the point-local potentials do not have massless
  limits, while the field strengths (see below) do.}
\end{Thm}
For spin one, our string-localized potential $\ASt_\mu(x,e)$ is fixed by
these requirements, see Prop.~\ref{AUnique}. There are no ghosts; all
fields are physical in the sense that they act in a Hilbert space and
their elementary excitations are the particles of the given mass and
spin. This also holds for the fields
$\phi^{(s,k)}_{\mu_1\cdots\mu_k}$, which we call {\em escort fields}.  

In the massless case, the construction of free string-localized
quantum fields via intertwiners has been performed for any
(half-integer) helicity by Plaschke and
Yngvason~\cite{PlaschkeYngvason}. Among these are potentials for the 
massless field strengths (see below). 
Interestingly, for spin $s\geq 2$ neither the
field strengths nor the string-localized potentials seem to coincide
with our massless limits. In the case of the field strength for $s=2$ (the
linearized Riemann tensor) this has been observed already
in~\cite{Veltman70}.     

We wish to emphasize one can construct an abundance of
string-localized fields with good UV behaviour ({\em c.f.\ }the
proof of Prop.~\ref{AUnique}), but without the
relation~\eqref{eqPotPotPts} to point-like fields they would be of no 
use for a perturbative construction along the lines sketched in the
Outlook. This is so since a ``properly string-local''   
interaction Lagrangean, {\em i.e.\ }not differing from a point-local one by a
divergence, would lead to completely de-localized interacting fields.  

Let us recall that for each spin $s$ there is a second well-known 
point-local quantum field acting in Hilbert space with optimal 
UV behaviour: the so-called field
strength tensor $\F_{\mu_1\nu_1\cdots \mu_s\nu_s}$.  
This rank-$2s$ tensor is symmetric under exchange of any of the pairs
$(\mu_i,\nu_i)\leftrightarrow (\mu_j,\nu_j)$ and anti-symmetric under 
exchange of any of the indices $\mu_i\leftrightarrow \nu_i$. 
It 
is related to the above-mentioned field
$\PotPt_{\mu_1\cdots \mu_2}$ by the PDE
\begin{align} \label{eqdAFs}
\F_{\mu_1\nu_1\cdots \mu_s\nu_s} &= 
\sum_{I\subset\{1,\ldots,s\}}(-1)^{|I|}
\partial_{\mu_{j_1}}\cdots \partial_{\mu_{j_{|I^c|}}}
\partial_{\nu_{i_1}}\cdots \partial_{\nu_{i_{|I|}}} 
\PotPt_{\nu_{j_1}\cdots \nu_{j_{|I^c|}}\mu_{i_1}\cdots \mu_{i_{|I|}}},
\end{align}
and this is why the quantum field $\PotPt_{\mu_1\cdots\mu_s}(x)$ 
is called ``tensor potential''. 
Our string-localized field $\Pot_{\mu_1\cdots\mu_s}(x,e)$ is also a tensor
potential in the sense that it satisfies Eq.~\eqref{eqdAFs} together
with $\PotPt_{\mu_1\cdots\mu_s}$. It comes about as an $s$-fold line
integral over the field strength, see Eq.~\eqref{eqPotIntegralF}
below. In fact, classically the properties of having the same field 
strength~\eqref{eqdAFs} and being related as in \eqref{eqPotPotPts} are
equivalent, in analogy to Poincar\'es Lemma (which states that a
closed form on $\RR^n$ is exact). Since we are not aware of a proof of
this analogue to Poincar\'e's Lemma for symmetric tensors, we prove it 
in Appendix~\ref{secPoincare}, see
Prop.~\ref{PoincareSym}.   
It is noteworthy that although the field strength tensor has scaling
dimension $s+1$, at least in the massless case its UV behaviour can be
considered as good as that of the scalar field, independent of the
helicity, if one requires so-called associate homogeneity as a 
renormalization condition~\cite{NST}. Generalizations of this idea to
the massive case can be found in~\cite{Nikolov16,VarillyGracia}. 

Let us recall the physical significance of the potentials for spin one
and two. 
For spin one, the potential $\APt_\mu$ is the Proca field, which could
physically describe the vector bosons of the weak interaction or, in
the massless limit, photons or gluons. 
Eq.~\eqref{eqPotPotPts} is reminiscent of, but should not
be confused with, a gauge transformation. 
In the non-Abelian cases (weak bosons or gluons), the field tensor
$F_{\mu\nu}=\partial_\mu \APt_\nu-\partial_\nu \APt_\mu$ would be the linear
approximation to the field strength. 
For spin two, the potential $\APt_{\mu\nu}\equiv \hpt_{\mu\nu}$
could, in the massless limit, describe the quantum fluctuations of the
metric field. 
The relation \eqref{eqdAFs} is then just the relation
between the linearized Riemann tensor $F_{\mu\nu\alpha\beta}\equiv 2 
R_{\mu\nu\alpha\beta}$ and the perturbation to a background metric in the linear 
approximation: 
\begin{equation} \label{eqRhPt'}  
R_{\mu\nu\alpha\beta} = 
\half\,\big( 
  \partial_{\mu} \partial_{\alpha} \hpt_{\nu\beta} 
 + \partial_{\nu} \partial_{\beta}  \hpt_{\mu\alpha}
- \partial_{\nu} \partial_{\alpha} \hpt_{\mu\beta} 
 - \partial_{\mu} \partial_{\beta}  \hpt_{\nu\alpha}  \,\big).
\end{equation}

The article is organized as follows. 
In Section~\ref{secFreeFieldInt} we recall the representation theory for
massive particles with spin and construct string-local ``Wigner
intertwiners'', that is, single particle wave functions with specific
intertwiner properties. We extend the argument of~\cite{MSY,MSY2} that
these objects are in a one-to-one correspondence with
string-localized free quantum fields for the given particle type. We
comment on the construction of string-localized fields as line
integrals over point-local 
fields, calculate their scaling degrees and give a criterion for the
existence of their massless limits.  
In Section~\ref{secSpinOne} we construct the string-localized vector
potential and its escort field for spin one.
We also compare our construction with the gauge-theoretic setting as
advocated by G.~Scharf and co-workers\cite{ScharfGauge}, and comment
on the massless limit. 
The spin-one Wigner intertwiners are the building blocks for the
higher spin fields, which we construct in Section~\ref{secSpinGen}. 
In each case, we calculate the two-point function and write the
string-localized potential and escort fields as line integrals
over point-like fields.   
Finally, we give an outlook on interacting models in Section~\ref{secFinal}.  
\section{Quantum fields and Wigner intertwiners} \label{secFreeFieldInt} 
Free fields are fixed by the single particle states which they create
from the vacuum.  Thus, the construction of free fields 
is reduced to the construction of certain ``intertwiners''
relating the single particle space with the target space of the field.  
(This point of view is elaborated in Weinberg's monograph~\cite{Weinberg}.) 
In this section, we shall establish the one-to-one correspondence
between free fields and intertwiners. All statements and formulas hold
for the string-local as well as for the point-local case (by simply
neglecting the variable $e$.)     

The state space of a bosonic particle with mass $m$ and spin $s\in\NN$ in  
four-dimensional Minkowski space carries an irreducible unitary
positive-energy representation $\Uirr$ of the proper orthochronous 
Poincar\'e group $\Po$. 
To set the stage, let us recall these representations. 
The spin characterizes an irreducible unitary representation $\Ds$ of 
the stabilizer subgroup in $\Lor$ of a reference momentum $\bar p$ on the
mass shell for $m>0$,    
$$ 
\Hyp \doteq \{ p \in\RR^4:\; p\cdot p=m^2,\,p^0>0 \}.    
$$
This subgroup is the group of rotations, $SO(3)$, and the
representation $\Ds$ acts in a Hilbert space $\HlittleS$ of dimension
$2s+1$, the so-called little Hilbert space. 
The representation $\Uirr$ of $\Po$ is induced by $\Ds$ as
follows. The representation space is $\Hirr:=L^2(\Hyp,d\mu;\HlittleS)$, where
$d\mu(p)$ 
is the Lorentz invariant measure on $\Hyp$, and $\Uirr$ acts
according to    
\begin{equation} \label{eqUirr}  
\big(\Uirr(a,\lor)\psi\big) (p) = e^{i p\cdot
a}\,\Ds(\WigRot(\lor,p))\,\psi(\lor^{-1}p)\,. 
\end{equation}
Here $\WigRot(\lor,p)\in SO(3)$ is the so--called Wigner rotation, defined by 
\begin{equation}\label{eqWigRot}  
 R(\lor,p) :=  B_p^{-1}\,\lor\;B_{\lor^{-1}p}, 
\end{equation} 
where $B_p$, $p\in\Hyp$, is a family of Lorentz transformations such
that $B_p: \bar{p}\mapsto p$.     
This representation extends to the full Poincar\'e group by adjoining
representers for the space reflection $P$ (or parity transformation)
and the time reflection $T\doteq -P$.   
The space reflection can be considered as an element of the stability
subgroup of our reference momentum $\bar p$ within the orthochronous
Lorentz group $\LorOrtho$, which is the full rotation group
$O(3)$. All integer spin representations $\Ds$ extend 
to $O(3)$, and so $\Uirr$ extends naturally to the parity
transformation by Eq.~\eqref{eqUirr}. To simplify things, one may 
choose the family  $B_p$, $p\in\Hyp$, so that 
\begin{equation}\label{eqPBpP} 
 PB_pP=B_{Pp}, 
\end{equation} 
which implies $\WigRot(P,p)=P$ and 
\begin{equation}\label{eqWigRotP} 
\WigRot(P\Lambda P,p) = P\, \WigRot(\Lambda,Pp)\, P.
\end{equation} 
Then the representer of the space reflection is given by  
\begin{align} \label{eqUP} 
 \big(\Uirr(P)\psi\big) (p) =  \Ds(P) \, \psi(Pp).
\end{align}
Similarly, one can adjoin an {\em anti}- unitary representer of the
time reflection to the representation of $O(3)$, that is, an
anti-unitary operator $\Ds(T)$  satisfying the relations
\begin{equation} \label{eqDsT}
\Ds(T)\,\Ds(R)\, \Ds(T)^{-1} = \Ds(TRT),\quad
\Ds(T)^{2}=\unity.  
\end{equation}
(Such an operator exists for every $s\in\NN$ and is unique up to
a factor, see Chapters~\ref{secSpinOne} and \ref{secSpinGen}.)   
One now defines an anti-unitary involution $\Uirr(T)$ by 
\begin{align} \label{eqUT} 
 \big(\Uirr(T)\psi\big) (p) :=  \Ds(T) \, \psi(-Tp).
\end{align}
Since the adjoint action of $T$ on the Lorentz group coincides with that of $P$,
Eq.~\eqref{eqWigRotP} implies that $\Uirr(T)$ in fact extends $\Uirr$
to a representation of the full Poincar\'e group. 
Note that the anti-unitary representer of the $\PT$ transformation
$PT\equiv -\unity$ is now given by  
\begin{align} \label{eqUPT} 
 \big(\Uirr(-\unity)\psi\big) (p) = \Ds(-\unity) \, \psi(p) ,
\end{align}
where $\Ds(-\unity)$ is the anti-unitary operator $\Ds(T)\Ds(P)$. 

We adopt the following convention for the components of (co-) vectors. 
Given a reference system,
{\em i.e.\ }a vierbein $\{e_{(0)},\ldots, e_{(3)}\}$ of orthonormal Lorentz
vectors (tangent vectors to Minkowski space), the contra-variant
components $\xi^\mu$ of a Lorentz vector $\xi$ are the expansion
coefficients in $\xi=\xi^\mu e_{(\mu)}$. The covariant components
$p_\mu$ of a momentum space vector $p$ are defined by $p_\mu\doteq p\cdot 
e_{(\mu)}$. As we have already done, we identify the space of Lorentz
vectors as well as its dual (momentum space) with $\RR^4$ via the
components, {\em e.g. } $p\mapsto (p_0,\cdots,p_3)$. The Lorentz
product then reads $p\cdot 
p=p_0^2-\|\bfp\|^2$ with $\|\bfp\|^2=p_1^2+p_2^2+p_3^2$. The
Lorentz invariant measure on the mass shell is  
\begin{equation} \label{eqdmum}
d\mu(p) \equiv d\mu_m(p) = \frac{d^3\bfp}{2\omega_m(\bfp)},\quad 
\omega_m(\bfp)\doteq (\|\bfp\|^2+m^2)^{\frac{1}{2}}.
\end{equation}
We choose the reference system such that the reference momentum $\bar
p$ is identified with $(m,\bs{0})$ in $\RR^4$. 
%
\subsection{From quantum fields to intertwiners} \label{Field-Int}
Suppose we are given a string-localized free quantum field. 
For the sake of generality, we consider here not only vector or tensor 
fields, but an $N$-component field $\field_r(x,e)$,
$r=1,\ldots,N$, which transforms in a covariant way under some matrix
representation $D$ of the Lorentz group, {\em
  i.e.\ }Eq.~\eqref{eqCovTens} is replaced by  
\begin{equation}  \label{eqFieldCov}
U(a,\Lambda)\, \field_r(x,e) \, U(a,\Lambda)^{-1} = \sum_{r'=1}^N
\field_{r'}(a+\Lambda x,\Lambda e) \; D(\Lambda)_{r'r}. 
\end{equation}
We shall also consider fields which are covariant under the parity
transformation $P:(x^0,\bs{x})\mapsto (x^0,-\bs{x})$, {\em i.e.}, the
  representations $U$ and $D$ extend to the orthochronous Lorentz group and
  Eq.~\eqref{eqFieldCov} also holds for $\Lambda=P$. 
The field is assumed to act in some Hilbert space which contains the
irreducible space $\Hirr$ for single particles of mass $m$ and spin $s$,
and also contains the vacuum vector $\Omega$, characterized up to a
factor by its invariance under Poincar\'e transformations.     
Suppose further, that the field creates from the vacuum single
particle states of mass $m$ and spin $s$.  
This means that the vector $\field_r(x,e)\Omega$ is in the single-particle space
$\Hirr$.\footnote{In the interacting case, it means that the above
  vector has a non-vanishing  projection into the single particle space.} 
The covariance
property~\eqref{eqFieldCov} with $\Lambda =\unity$ then implies 
that there is a family of $\HlittleS$-valued distributions $\IntFct_r(p,e)$,
$r=1,\ldots, N$, such that for $p\in\Hyp$   
\begin{equation} \label{eqIntFct}
\big(\field_r(x,e)\Omega\big)(p) \doteq (2\pi)^{-\frac{3}{2}}\,
e^{ip\cdot x}\, \IntFct_r(p,e)\quad 
\in\; \HlittleS. 
\end{equation}
We call $\IntFct_r(p,e)$ the {\em intertwiner function} of the field
$\field_r$. (The factor $(2\pi)^{-\frac{3}{2}}$ has been introduced in
order to have canonical commutation relation relations in the scalar
case if one chooses $\IntFct\equiv 1$.)  
We consider here the case when the antiparticle coincides with
the particle: The adjoint of $\field_r(x,e)$ creates particles of the
same type from the vacuum, that is, $\field_r(x,e)^*\Omega$ is also in
$\Hirr$. Thus, there is a $\HlittleS$-valued distribution
$\IntFctC_r(p,e)$, the so-called conjugate intertwiner, such that for
$p\in\Hyp$    
\begin{equation} \label{eqIntFctC}
\big(\field_r(x,e)^*\Omega\big)(p) \doteq  (2\pi)^{-\frac{3}{2}}\, 
e^{ip\cdot x}\, \IntFctC_r(p,e)\quad
\in\; \HlittleS.  
\end{equation}
We emphasize that all formulas have to be understood in the sense of
distributions, {\em i.e.\ }after smearing in $x$ and $e$:  
Let us spell out for later reference the rigorous sense of
Eq.~\eqref{eqIntFct}. Given a pair of test functions
$f\in\calS(\RR^4)$ and $h\in\calD(\Spd)$, let $\field_r(f,h)$ be the smeared
field\footnote{Of course rigorously speaking, this object is first
  defined and $\field(x,e)$ is informally defined by
  Eq.~\eqref{eqSmearedField}.}     
\begin{equation} \label{eqSmearedField}
\field_r(f,h) =  \int d\sigma(e)\, h(e) \int d^4 x\, f(x)\,
\field_r(x,e),   
\end{equation}
where $d\sigma(e)$ is the Lorentz invariant measure on
$\Spd$,\footnote{The measure $d\sigma$ is given as follows. After
  choosing a Lorentz reference system, a point $e\in\Spd$ corresponds
  to a point $(e^0,\bs{e})\in\RR\times\RR^{3}$, and the spatial part
  $\bs{e}$ must be of the form $(1+(e^0)^2)^{1/2}\bs{n}$, where
  $\bs{n}$ is in the unit sphere $S^{2}\subset \RR^{3}$. Then 
\begin{equation}   \label{eqMeasuredS}
\int_\Spd d\sigma(e) h(e) = \int_\RR de^0  \,\big(1+(e^0)^2\big) \int_{S^{2}}
d\Omega(\bs{n})\, h(e^0,\bs{n}), 
\end{equation}
where $d\Omega^2$ is the metric on the unit sphere $S^{2}$.
}
and let $\IntFct_r(p,h)$ be the intertwiner function smeared in $e$,   
\begin{equation}  
\IntFct_r(p,h)  \doteq \int d\sigma(e)\, h(e) \,\IntFct_r(p,e)  . 
\end{equation}
Then Eq.~\eqref{eqIntFct} means that the single particle vector
$\field_r(f,h)\Omega$, viewed as a $\HlittleS$-valued function 
on the mass shell, is given by 
\begin{equation} \label{eqIntFctSmeared} 
\big(\field_r(f,h)\Omega\big)(p) = (2\pi)^{-\frac{3}{2}}\,\hat
f(p)\,\IntFct_r(p,h),
\end{equation}
where $\hat f$ is 
the Fourier transform of $f$.\footnote{We adopt the following convention for
  the Fourier transform $\hat f$, whose inverse we denote by $\check{f}$:  
\begin{equation}  \label{eqFT}
\hat f(p)\doteq \int d^4x \, e^{ip\cdot x} f(x), \quad 
\check{f}(x) = (2\pi)^{-4} \int d^4p \, e^{-ip\cdot x} f(p). 
\end{equation}
} 

The next Lemma states the specific properties of the intertwiner function
which are implied by covariance and locality of the field. To
formulate it, we interpret the family $\IntFct_r(p,e)$ as a linear map
from some $N$-dimensional vector space $\target$ with basis
$\{e_{(1)},\ldots,e_{(N)}\}$, the ``target space'' of the field, to
$\HlittleS$ via   
\begin{equation} \label{eqvre}
v(p,e)\, e_{(r)} \doteq v_r(p,e) \qquad \in\HlittleS ,   
\end{equation}
and the matrix $D(\Lambda)_{r'r}$ as a linear endomorphism of $\target$
in the usual way. Further, we denote by $\Spdc$ the complexification
of $\Spd$, 
\begin{align} \label{eqSpdc}
\Spdc  & :=  \{ \spdc \in\CC^4,\; \spd\cdot \spd  = -1  \},   
\end{align} 
where the dot is the bi-linear extension of the Minkowski metric to
$\CC^4$,   
and by $\Tub$ the tuboid consisting of all points in $\Spdc$ whose
imaginary part is in the interior of the forward light cone
$V_+\subset \RR^4$. 
We will consider subsets $\Theta$ of $\Tub$ of the form 
\begin{equation} \label{eqTheta}
\Theta= \Spdc\cap \big(K_1+i\RR^+K_2\big),
\end{equation} 
where $K_1$ and $K_2$ are compact subsets of $\RR^4$ and $K_2$ is
contained in the  forward light cone. 
\begin{Prop}  \label{Int}
$i)$ The intertwiner function $\IntFct$ satisfies the {\em
  intertwiner relation}:  
\begin{align} \label{eqInt}
\Ds\big(R(\Lambda,p)\big) \circ \IntFct(\Lambda^{-1}p,\Lambda^{-1}e) 
& = \IntFct(p,e)\circ D(\Lambda), \quad \Lambda\in\Lor. 
\end{align}
The conjugate intertwiner $\IntFctC$ satisfies the same relation, but
with $D(\Lambda)$ replaced by the component-wise complex conjugate
$\overline{D(\Lambda)}$. 
If the field is parity covariant, then these relations are satisfied for all
$\Lambda\in L^\uparrow$.  

$ii)$ For almost all $p$, the distribution $\IntFct_r(p,e)$ is the
boundary value of a function
$\tilde\IntFct_r(p,e)$\footnote{\label{DistBdValue}%
That means that for
  $h\in\calD(\Spd)$ and almost all $p$, $\IntFct_r(p,h)$  is obtained as a
  (weak) integral   
\begin{align} \label{eqIntDist} 
\IntFct_r(p,h) = \int_\Spd d\sigma(e) \,h(e) \, \tilde \IntFct_r(p,e),  
\end{align}
where one lets the argument $\spd$ of $\tilde\IntFct_r(p,e)$ approach  
$\Spd$ from $\Spdc$ inside the tuboid $\Tub$ after the
integration, see \cite[Thm.\ A.2]{BrosMos}. 
} 
on $\Hyp\times
\Tub$ which is, 
for almost all $p$, analytic in $e\in\Tub$ and satisfies the following bound: 
There is a constant $n\in\NN_0$ and a function $M$ on $\Hyp$ 
which is locally $L^2$ w.r.t.\ $d\mu$ and polynomially bounded,   
and for every $\Theta\subset\Tub$ of the form indicated in
Eq.~\eqref{eqTheta}, there is a constant $c=c_\Theta$ such that for
all $e\in\Theta$ holds 
\begin{equation} \label{eqModGrowth}
\|\tilde\IntFct_r(p,e)\| \leq c \, M(p) \, |\Im e|^{-n}. 
\end{equation} 
Here, $|\cdot|$ denotes any norm in $\RR^{4}$ and the norm of
$\tilde\IntFct_r$ refers to the little Hilbert space $\HlittleS$.    

$iii)$ The conjugate intertwiner is uniquely fixed by $\IntFct$ via the
relation\footnote{In odd-dimensional space-time, the PT
  transformation $-\unity$ must be replaced by the reflection $j_1$ at the edge
  of a wedge, see Eq.~\eqref{eqW1} below, and Eq.~\eqref{eqIntIntC} must be
  replaced by Eq.~\eqref{eqIntIntCJ}.  
}
\begin{equation}  \label{eqIntIntC}
\Ds(-\unity) \,\IntFctC(p,e) \, e_{(r)}  = \IntFct(p,-e)\,
D(-\unity) \, e_{(r)}. 
\end{equation}
Here, $\D(-\unity)$ arises from the unit component via analytic
continuation through the complex proper Lorentz group
$L_+(C)$.\footnote{Note that all finite dimensional representations
  $\D$ extend analytically to the complex Lorentz group $L_+(C)$, and
  that the latter is path connected. A path from $\unity$ to $-\unity$
  is for example $\Lambda_1(is)$, $s\in[0,\pi]$, composed with $R_1(\alpha)$,
  $\alpha\in[0,\pi]$.}    
\end{Prop}
Note that  $\D(-\unity)$ is linear, while $\Ds(-\unity)$ is anti-linear.

If the field $\field_r$ is interacting, its intertwiner function
$\IntFct_r$ can defined in the same way, but with the projection
operator onto $\Hirr$ on the left hand side of Eq.~\eqref{eqIntFct},
and it has the same intertwiner and analyticity properties as stated
in the Lemma.   
\begin{Proof}
Applying Eq.~\eqref{eqFieldCov} with $a=0$, one finds that the
family of distributions $\IntFct_r$ must satisfy 
\begin{equation*} 
\Ds\big(R(\Lambda,p)\big) \IntFct_r(\Lambda^{-1}p,\Lambda^{-1}e) 
= \sum_{r'=1}^N \IntFct_{r'}(p,e)\, D(\Lambda)_{r' r}.
\end{equation*}
This yields Eq.~\eqref{eqInt}. A similar argument holds for the
conjugate intertwiner. 

Part $ii)$ and $iii)$ have been shown in \cite[Thm.\ 3.3
  $(iii)$]{MSY2} for the trivial representation $D(\Lambda)=1$, using
the Bisognano-Wichmann property, which has been proved to hold in
massive string-localized models~\cite{M01a}. One considers the wedge
$W_1$, 
\begin{equation} \label{eqW1} 
W_1 \doteq \left\{  x\in\RR^4|\; x^1 > \left|  x^{0}\right|  \right\},  
\end{equation}
together with the one-parameter group $\Lambda_1(\cdot)$ of Lorentz boosts 
which leave $W_1$ invariant, and the reflection $j_1$ across the
edge of the wedge. 
The latter results from analytic extension of the (entire analytic)
$L_+(C)$-valued function $\Lambda_1(z)$ at $z=i\pi$, and also can be
represented (in 4 dimensions) as the composition of the \PT {}
transformation with a $\pi$-rotation about the $1$-axis, $j_1 = - R_1(\pi)$. 
Choosing the family  $B_p$, $p\in\Hyp$, so that 
\begin{equation} \label{eqRBp} 
R_1(\pi) \circ B_p = B_{R_1(\pi) p}\circ R_1(\pi),
\end{equation} 
the Wigner rotation satisfies $\WigRot(R_1(\pi),p)=R_1(\pi)$ for every $p$, and
one gets, using Eq.~\eqref{eqUT}: 
\begin{align} \label{eqUj} 
 \big(\Uirr(j_1)\psi\big) (p)  = \Ds(j_1) \, \psi(-j_1 p).
\end{align}
(This is the same representer as in \cite{MSY2}.) From here, we have
the same assumptions as \cite{MSY2}, except that our $\D$ is not
trivial. However, it is a finite dimensional representation which
extends analytically to the complex Lorentz group, so all arguments
concerning analyticity and bounds go through.   
The only modification is the relation between $\IntFctC$ and
$\IntFct$, 
which corresponds to Eq.~(30) in \cite{MSY2}. Instead of the latter, we
arrive at 
\begin{equation}  \label{eqIntIntCJ}
\Ds(j_1) \,\IntFctC(-j_1p,e) \, e_{(r)} = \IntFct(p,j_1e)\,
D(j_1) \, e_{(r)},  
\end{equation}
where the right hand side is defined as the analytic extension of 
$\IntFct(p,\Lambda_1(t)e) D(\Lambda_1(t)) e_{(r)}$ at $t=i\pi$ (after
smearing in $e$).   
This implies the claimed relation~\eqref{eqIntIntC} by applying the
intertwiner properties of $\IntFct$ and $\IntFctC$ to the rotation $R_1(\pi)$. 
\end{Proof}
\begin{Def}[Intertwiners] \label{DefIntFct}
A family of distributions $\IntFct_r$, $r=1,\ldots,N$,  with the 
intertwiner $(i)$ and analyticity $(ii)$ properties of Prop.~\ref{Int}
is called a {\em Wigner intertwiner} from $D$ to $\Ds$.  
Given a Wigner intertwiner $\IntFct$, its {\em conjugate} $\IntFctC$
is defined by Eq.~\eqref{eqIntIntC}. $\IntFct$ is called {\em
  self-conjugate} if $\IntFct=\IntFctC$.  
\end{Def}

Note that for the reference momentum $p=\bar p$, the intertwiner
$\IntFct(\bar p,e)\doteq \hat\IntFct(e)$  
satisfies  the ``small intertwiner relation'' 
\begin{equation} \label{eqIntInt}
\Ds(R) \circ \hat\IntFct(R^{-1}e) = \hat\IntFct(e)\circ D(R)
,\quad R\in SO(3), 
\end{equation}
since for $R\in SO(3)$ the Wigner rotation $\WigRot(R,\bar p)$
coincides with $R$.    
Conversely, the intertwiner function can be recovered from 
$\hat{\IntFct}(e)$ by the identity  
\begin{equation} \label{eqIntInt'}
\IntFct(p,e) = \hat\IntFct(B_p^{-1}e) \circ D(B_p^{-1}),   
\end{equation}
which also follows from the intertwiner relation~\eqref{eqInt} since
$\WigRot(B_p,p)=\unity$.  
Thus, the construction of intertwiner functions boils down to finding
solutions to the relation \eqref{eqIntInt}.
\subsection{Via intertwiners to free fields} \label{Int-Field}
Conversely, one can construct a free field via Wigner intertwiners, as
is done in Weinberg's monograph~\cite{Weinberg} for point-local
fields. The first step is to construct a Wigner intertwiner from a
given representation $\D$ of the Lorentz group to $\Ds$. This may be 
accomplished as follows.   
First find a solution $\hat{\IntFct}(e)$ to the ``small
intertwiner relation''~\eqref{eqIntInt}, which has the required
analyticity properties, namely, which is the boundary value of an
analytic function on the tuboid $\calT_+$. Then define $\IntFct(p,e)$ as in
Eq.~\eqref{eqIntInt'}. This is a Wigner intertwiner in the sense
of Definition~\ref{DefIntFct}. From here, define
the conjugate intertwiner $\IntFctC(p,e)$ by Eq.~\eqref{eqIntIntC}. 
 
In a second step, one constructs a free field from the Wigner
intertwiner and its conjugate. 
We denote by $\calH$, $U$ and $\Omega$ the bosonic Fock space over
$\Hirr$, the second quantization of the single particle representation
$\Uirr$, and the invariant Fock space vacuum, respectively.  
Let further denote $a^*(\psi)$ and $a(\psi)$,  $\psi\in\Hirr$, the
creation and annihilation operators.  
Given a pair of test functions $f\in\calS(\RR^4)$ and
$h\in\calD(\Spd)$, let $\psi_r$ 
be the single particle vector defined by the right hand side of
Eq.~\eqref{eqIntFctSmeared}, and analogously let $\psi_r^c$ 
be defined by 
$$
 \psi_r^c(p) \doteq
 (2\pi)^{-\frac{3}{2}}\,\hat{\bar{f}}(p)\,\IntFctC_r(p,\bar h).
$$ 
(Of course, $\psi_r$ depends linearly on the test functions $(f,h)$,
 while $\psi_r^c$ depends anti-linearly on them.)  Then define 
$$
\field_r(f,h)\doteq a^*(\psi_r)+a(\psi_r^c). 
$$
By a version of the Jost-Schroer theorem~\cite{SW,
  St,Mu_JoSch}, this is the unique (up to unitary equivalence) free
field with the given intertwiner function $\IntFct$. 
Let us re-write it in the usual informal notation, writing symbolically 
$$
a^*(\psi) =: \int_{\Hyp} d\mu(p)\sum_{k=1}^{2s+1} \psi^k(p) a^*(p,k),\qquad 
a(\psi)  =: \int_{\Hyp} d\mu(p)\sum_{k=1}^{2s+1}
d\mu(p)\,\overline{\psi^k(p)} a(p,k). 
$$
Here, the subscript $k$ denotes the components with respect to a 
basis $\{e_{(k)}\}$ in $\HlittleS$, i.e., $\psi(p) = \sum_k
\psi^k(p) \, e_{(k)}$.  
Then 
\begin{equation} \label{eqFreeField}
\field_r(x,\spd) =  
(2\pi)^{-\frac{3}{2}}\,\int_{\Hyp}\d\mu(p) \sum_{\sz=1}^{2s+1} \left\{  \;
e^{ip\cdot x}  \; \IntFct_r^k(p,e) \, a^*(p,\sz) 
+ e^{-ip\cdot x} \; \overline{{\IntFctC_r}^{,k}(p,e)} \,a(p,\sz)\;
\right\}\,. 
\end{equation}
We call this {\em the free field for} the Wigner intertwiner
$\IntFct$. Of course, it is hermitean if and only if the intertwiner is
self-conjugate, $\IntFctC=\IntFct$.  
The two-point function of two such fields $\field_{1,r}$ and
$\field_{2,r}$ with respective Wigner intertwiners $\IntFct_1$ and
$\IntFct_2$ comes out as  
\begin{align}   \label{eq2PtGeneral}
\lsp \Omega, \field_{1,r}(x,\spd)\,\field_{2,r'}(x',\spd')\Omega\rsp 
=& (2\pi)^{-3}\,\int_{\Hyp} d\mu(p)\; e^{-ip\cdot  (x-x')}\,
\MTwoPt_{r,r'}^{\field_1\field_2}(p,e,e')\, \\
\MTwoPt_{r,r'}^{\field_1\field_2}(p,e,e')= & \lsp
\IntFct_{1r}^c(p,e),\IntFct_{2r'}(p,e')\rsp_{\HlittleS}, 
\label{eqM}
\end{align}
where $\lsp \cdot, \cdot \rsp$ and $\lsp \cdot , \cdot
\rsp_{\HlittleS}$  denote the scalar products in Fock space $\calH$
and in the little Hilbert space $\HlittleS$, respectively. 
The distribution $M_{r,r'}^{\field_1\field_2}(p,e,e')$ is the Fourier
transform of the two-point function after splitting off the mass shell
delta contribution, and shall be called the {\em on-shell part of the two-point
  function}. 
Note that positivity of the two-point function 
is satisfied by construction.  
\begin{Prop}  \label{Locality}
Let $\IntFct(p,e)$ be a Wigner intertwiner from $\Ds$ to $\D$ in the
sense of Definition~\ref{DefIntFct}, and let $\IntFctC(p,e)$ be
defined by Eq.~\eqref{eqIntIntC}. Then the 
field defined in Eq.~\eqref{eqFreeField} is an operator valued
distribution.\footnote{\label{OpValDist}By this we mean that its
  vacuum expectation values 
  are distributions~\cite{SW}.}
It is string-localized and covariant in the sense of Eqs.~\eqref{eqFieldLoc} and
\eqref{eqFieldCov}. 
It further satisfies the Reeh-Schlieder and Bisognano-Wichmann
properties (see below).  
Moreover, the CPT symmetry holds:  
\begin{equation}  \label{eqCPT}
 U(-\unity) \, \field_r(x,e)\,U(-\unity)^{-1}= \sum_{r'=1}^N
 \field_{r'}(-x,-e)^* \D(-1)_{r'r}. 
\end{equation}
Fields constructed via~\eqref{eqFreeField} by different Wigner
intertwiners from $\Ds$ to the same $\D$ are relatively string-local
to each other.  
\end{Prop}
Needless to say, if the Wigner intertwiner satisfies the intertwiner
relation~\eqref{eqInt} also for the parity transformation, then the
field is parity covariant.  

By the {\it Reeh-Schlieder property} we mean that products of the
fields already generate a dense set from the vacuum when smeared
within arbitrary, fixed, open sets $\calO\in\RR^4$ and $U\in\Spd$.  
The {\it Bisognano-Wichmann property} means that the modular group
 and modular conjugation of the algebra associated to a wedge $W$
 coincides with the representers of the boosts $\Boo{W}{t}$ and
 reflection $j_W$ associated to $W$, respectively, see e.g.~\cite{H96}.  
\begin{Proof}
To show the (weak) distribution property, it suffices by Wick's
theorem to show that $\field_r(x,e)\Omega$ is a distribution with 
values in the single particle space. To this end, note that Eq.~(A.6)
of Lemma~A.3 in~\cite{MSY} implies that $\IntFct_r(p,h)$, as defined
in Eq.~\eqref{eqIntDist}, is bounded in norm by
$$ 
\|\IntFct_r(p,h)\|_{\HlittleS} \leq M(p) \, N(h),
$$
where $M$ is a positive function on the mass shell which is locally
$L^2$ w.r.t.\ $d\mu(p)$ and polynomially bounded, 
and $N(h)$ is a sum of semi-norms from the topology of
$\calD(\Spd)$. Writing out the smeared vector $\field_r(f,h)\Omega$ as in
Eq.~\eqref{eqIntFctSmeared}, this yields the bound 
$$
\|\field_r(f,h)\Omega\|^2 \leq N(h)^2 N'(\hat f)^2 (2\pi)^{-3}\int
d\mu(p) (1+|p|^4)^{-1} M(p)^2 \leq c \, N(h)^2 N''(f)^2,  
$$
where $N'(\cdot)$ and $N''(\cdot)$ are (weighted) sums of semi-norms  in
the Schwartz-topology, and $|p|$ is some euclidean norm of $p$. (We
have chosen $N'$ so that $|\hat f(p)| \leq (1+ |p|^4)^{-1} N'(\hat f)$, and used 
that there exists $N''(f)$ dominating  $N'(\hat f)$, by
continuity of the Fourier transform.) This concludes the proof that
the field is a (weak) distribution. 

The field is defined just so that Eqs.~\eqref{eqIntFct} and 
\eqref{eqIntFctC} hold.  
Thus the single particle vector $U(\Lambda)\field_r(x,e)\Omega$ is, by 
construction, given by the
following $\HlittleS$-valued function on the mass shell:  
$$
\big(U(\Lambda)\field_r(x,e)\Omega\big)(p)=e^{i\Lambda ^{-1}p\cdot x}
\Ds(R(\Lambda,p))\,v_r(\Lambda^{-1}p,e).
$$
By the intertwiner property~\eqref{eqInt} of $\IntFct(p,e)$ the right
hand side coincides with 
$$
e^{ip\cdot \Lambda x} \sum_{r'=1}^N v_{r'}(p,\Lambda e)\D(\Lambda)_{r' r} \equiv 
\sum_{r'=1}^N \big(\field_{r'}(\Lambda x, \Lambda e)\Omega\big)(p)\,
\D(\Lambda)_{r' r}.  
$$ 
Since $\Omega$ is invariant under $U$, we have shown that the
identity~\eqref{eqFieldCov} of operators holds if applied to the
vacuum vector.  
By the combinatorial arguments used in the Jost-Schroer theorem (Wick
expansion of the $n$-point functions), this implies that the operators
coincide, hence Eq.~\eqref{eqFieldCov}.  
A similar argument shows the CPT symmetry~\eqref{eqCPT}. 
 
Since the proof of string-locality~\eqref{eqFieldLoc} in \cite{MSY2}
relies on the concept of modular localization which we don't explain
here, we give a direct proof of relative string-locality, in a slight
adaption of the proof in \cite{MSY}. 
Let $\IntFct'(p,e)$ be a second Wigner intertwiners from $\Ds$ to $\D$
and let $\field'_r(x,e)$ be the corresponding field.  
Suppose, $(x,e)$ and $(x',e')$ satisfy the condition for 
Eq.~\eqref{eqFieldLoc}. Then $\String_{x,e}$ and $\String_{x',e'}$ are
space-like separated {\em and} $e$ and $e'$ are also space-like
separated~\cite[Lemma~A1]{MSY2}. 
This implies that there is a wedge region $W$ such that $x,e\in W$ and
$x',e'\in W'$, see~\cite{BGL,MSY2}, where $W'$ denotes the causal
complement of $W$.    
Let $j_W$ and $\Lambda_W(t)$ be the reflection and the boosts,
respectively, corresponding to $W$, {\em i.e.},
$j_W\doteq (a,\Lambda) \,j_1 \, (a,\Lambda)^{-1}$ and $\Lambda_W(t)\doteq
(a,\Lambda) \Lambda_1(t) (a,\Lambda)^{-1}$ if $W=(a,\Lambda) W_1\equiv a+\Lambda
W_1$.  Denote by $g_t$ the proper non-orthochronous Poincar\'e transformation
$\Lambda_W(-t)j_W$. Then one verifies the relation  
\begin{multline}
\lsp \Omega,\field_r(x,e)^* \field'_{r'}(g_t^{-1} x',g_t^{-1} e') \Omega \rsp   
\\ =\sum_{s,s'}\lsp \Omega,\field'_{s'}(x',e') \field_s(g_t x,g_t e)^*
\Omega \rsp\; \overline{D(g_t)_{sr}}\,
D(g_t)_{s'r'}. \label{eqLocTwoPt} 
\end{multline} 
(We have successively used invariance of $\Omega$ under $U\equiv
U(g_t)$,  anti-unitarity of $U$, namely $\lsp \Omega,\psi\rsp =\lsp
U^{-1}\Omega,\psi\rsp = \lsp U\psi,\Omega \rsp$, and then
covariance~\eqref{eqFieldCov} and the CPT
symmetry~\eqref{eqCPT}. Finally we have adjoined the field operators
to the right hand side of the scalar product.) The matrix-valued
function  $\D(g_t)$ (and hence $\overline{\D(g_{\overline{t}})}$) is
entire analytic in the boost parameter $t$. Note that $j_W$ and
$\Lambda_W(t)$ commute, hence $g_t^{-1}=g_{-t}$, and that for $t$ in
the strip $\RR+i(0,\pi)$ the imaginary parts of $g_tx$, $g_te$,
$g_{-t}x'$ and $g_{-t}e'$ all lie in the forward light cone $V_+$ (see
for example Eq.~(A.7) in~\cite{MSY2}). Now the two-point function is
an analytic function of the second $x$-variable in the tube
$\RR^4+iV_+$ due to the support of its Fourier transform, and also of
the second $e$-variable in the tuboid $\calT_+$ due to the analyticity
of the intertwiner function.  Therefore Eq.~\eqref{eqLocTwoPt}
extends, by the Schwarz 
reflection principle, from the boundary at $\Im t=0$ over the entire
strip to the upper boundary, and the boundary values at $t=i\pi$
coincide for both sides. But $\Lambda_W(\pm i\pi)=j_W$, i.e., $g_{\pm
  i\pi}=\unit$, and thus Eq.~\eqref{eqLocTwoPt} at $t=i\pi$ is just
the locality of the two-point functions. This implies locality of the
fields by the usual Jost-Schroer arguments.  

The Reeh-Schlieder and Bisognano-Wichmann properties are shown as in
\cite[Thm.\ 3.3]{MSY2}. 
\end{Proof}
\paragraph{Remark on the point-local case.}  
The intertwiner concept allows for an easy proof that a free point-local
field for a given particle type $(m,s)$ and transforming under an {\em
  irreducible} representation $D$ is {\em unique} up to unitary
equivalence: 
In the point-local case, the intertwiner does not depend on
$e$. Hence the map $\hat\IntFct\equiv \IntFct(\bar p)$ from $\target$
to $\HlittleS$ is also $e$-independent, and Eq.~\eqref{eqIntInt} means
just that it is an intertwiner in the usual sense. Such intertwiner exists if 
and only if $D|SO(3)$ contains $\Ds$ as a sub-representation. 
This is the case iff $D$ contains some irreducible representation
$D^{(j,k)}$ with $|j-k|\leq s \leq j+k$. Each such representation
$D^{(j,k)}|SO(3)$ contains  
the spin-$s$ representation of the rotation group exactly 
once, and hence the corresponding intertwiner $\hat\IntFct^{(j,k)}$ is unique
up to a factor. Consequently, if the representation $D$ is {\em
  irreducible}, then the field is unique (up to unitary
equivalence). 

This uniqueness  does not hold for 
string-localized fields! In particular, $\IntFct(p,e)$ 
can be multiplied with $F(p\cdot e)$, where $F$ is the boundary value
of any (numerical) analytic function on the upper complex half plane.  
\subsection{Line integrals}  \label{secLineDef} 
All string-localized fields we are treating in this article can be
viewed as $n$-fold 
line integrals over point fields. These integrals are to be understood in
the following sense. Let $n\in\NN$ and let  $\field^\pt(x)$ be some
free point-local field. 
(It may be a tensor field, but we omit the tensor indices). 
Then for fixed $e\in\Spd$ and $\nu\in\NN$ the integral 
$$
\field_{(\nu)}(x,e)\doteq \int_0^\nu ds_1\cdots \int_0^\nu ds_n \,
\field^\pt\big(x+(s_1+\cdots + s_n)e\big) 
$$
exists as a distribution in $x$, since smearing it with a test
function $f(x)$ is the same as smearing  $\field^\pt(x)$ with the test
function   
$$
x\mapsto \int_0^\nu  ds_1\cdots \int_0^\nu ds_n \, f\big(x-(s_1+\cdots
+ s_n)e\big).  
$$
However, in the limit $\nu\to \infty$, this is not a Schwartz function
any more.   
Then we have to smear also in $e$, i.e., we have to understand
$\field_{(\nu)}(x,e)$ as a distribution in $x$ {\em and} $e$,
that is, as a string-localized field. 
The question is if the two limits in the single particle space  
\begin{equation} \label{eqSinglePartIntegral} 
\psi\doteq \lim_{\nu\to \infty} \field_{(\nu)}(f,h)\Omega,\qquad
\psi^{c} \doteq \lim_{\nu\to \infty} \field_{(\nu)}(f,h)^*\Omega  
\end{equation}
exist. (String-locality~\eqref{eqFieldLoc} and
covariance~\eqref{eqFieldCov} are then readily verified.) One finds
that the intertwiner function of the string-localized  
field $\field_{(\nu)}(x,e)$ and its conjugate are given by 
\begin{equation} \label{eqvvcSt}
\IntFct_\nu(p,e) = t_\nu^n(p,e) \, \IntFctPt(p) ,  \qquad
\IntFct_\nu^{c}(p,e) = t_\nu^n(p,e) \, \IntFctPtC(p),   
\end{equation}
where $\IntFctPt$ and $\IntFctPtC$ are the intertwiner function, and
its conjugate, of the point-local field $\field^\pt(x)$, and where
$t_\nu^n(p,\cdot)$ is, for $p\in\Hyp$, the following distribution on $\Spd$: 
\begin{equation}   
t_\nu^n(p,e)\doteq \int_0^\nu ds_1\cdots \int_0^\nu ds_n\, 
e^{i(s_1+\cdots +s_n)p\cdot e}.  
\end{equation}
In order to show that the limits in Eq.~\eqref{eqSinglePartIntegral}
exist in the $L^2$-norm of the single particle space, we need a
uniform bound in $p$, which we provide in the next Lemma. 
\begin{Lem} \label{Dist1/pe}
For $h\in \calD(\Spd)$ and $p\in\Hyp$, the limit $\lim_{\nu\to\infty}
t_\nu^n(p,h)$ exists and is given by 
\begin{equation} \label{eqtph}  
 t^n(p,h)\doteq  \lim_{\eps\to 0} 
\int_\Spd d\sigma(e)\,  h(e) \, \frac{i^n}{(p\cdot e+i\eps)^n}.
\end{equation}
Further, for every  for $h\in \calD(\Spd)$ there is a constant $N$
such that for all $p\in\Hyp$ the following bound holds, uniformly in $\nu$: 
\begin{equation}  \label{eqtphBd}
|t_\nu^n(p,h)| \leq \frac{N}{(p_0)^n}. 
\end{equation}
\end{Lem}
(Of course, the same bound holds for the limit $t^n(p,h)$.) 
\begin{Proof}
Note that the smeared distribution $t_{\nu}^n(p,h)$ can be written as 
\begin{equation} \label{eqtnuhtilde}
t_\nu^n(p,h) = \int_0^\nu ds_1\cdots \int_0^\nu ds_n\,
\tilde{h}\big((s_1+\cdots +s_n)p\big), 
\end{equation}
where 
\begin{equation} \label{eqhtilde} 
\tilde{h}(p)\doteq \int_\Spd d\sigma(e)\, h(e) \, e^{ip\cdot e}
\end{equation}
is the Fourier transform of the distribution $h(e)\delta^{(4)}(e\cdot
e+1)$. This is a 
distribution in $\RR^4$ with compact support, so its Fourier transform
is a smooth function, and decreases rapidly outside the wave front set
of the distribution $\delta^{(4)}(e\cdot e+1)$, 
see~\cite{Hormander}. The latter is just the set of co-vectors $(e,k)$
such that $k$ is orthogonal to the tangent space to $\Spd$ at
$e$~\cite[Example 8.2.5]{Hormander}, which is
$e^\perp$~\cite{ONeill}. That is to say, the wave front set 
of $\delta^{(4)}(e\cdot e+1)$ is the set of co-vectors $(e,k)$ with
$e\cdot e=-1$ and $k\in\RR e$. 
In particular, the wave front set contains no time-like co-vectors, hence 
$\tilde h(p)$ is of rapid decrease within the forward light cone. 
(This fact can also be shown directly, see Lemma~\ref{htilde}.)   
Thus, one sees from Eq.~\eqref{eqtnuhtilde} that the limit  of
$t_\nu^n(p,h)$ for $\nu\to \infty$ exists for $p\in\Hyp$. Writing 
$$
t_\nu^n(p,h) = \frac{1}{(p_0)^n} 
\int_0^{\nu p_0} ds_1\cdots \int_0^{\nu p_0} ds_n\,
\tilde{h}\big((s_1+\cdots +s_n)p/p_0\big),    
$$ 
and observing that the euclidean norm of the co-vector
$p/p_0=(1,\frac{\bfp}{p_0})$ is larger than one, 
we find the uniform bound for $t_\nu^n(p,h)$ in Eq.~\eqref{eqtphBd}.  
Finally, the identity of distributions 
$$
\lim_{\nu\to\infty} \int_0^\nu ds \, 
e^{i s p\cdot e} = \lim_{\eps\to 0}
\frac{i}{p\cdot e +i\eps} 
$$ 
on $\Spd$ (for fixed $p\in\Hyp$) yields Eq.~\eqref{eqtph}. 
This identity follows from the well-known fact that $\int_0^\infty ds\,
e^{is \omega }$ is just the Fourier transform of the  
Heaviside distribution, namely, $\frac{i}{\omega+i\eps}$, taking into
account that its pull-back under the map $e\mapsto p\cdot e$ is
well-defined~\cite[Thm.~8.2.4]{Hormander} since this map is a coordinate
function on $\Spd$.  This completes the proof. 
\end{Proof}
These facts imply that the two limits~\eqref{eqSinglePartIntegral} 
exist in the single particle space $L^2(\Hyp,\HlittleS)$, namely, 
$$
\psi(p) =  (2\pi)^{-\frac{3}{2}}\,\hat f(p) \,t^n(p,h) \,\IntFctPt(p)
,\quad \psi^c(p) = (2\pi)^{-\frac{3}{2}}\,\hat{\bar{f}}(p) \,t^n(p,\bar
h) \,\IntFctPtC(p).  
$$ 
This implies in turn that $\field_{(\nu)}(f,h)$ converges weakly to $a^*(\psi) +
a(\psi^c)$ in the sense of matrix elements between vectors with finite
particle number. Then the limit $\field(f,h) \doteq a^*(\psi) + a(\psi^c)$ 
defines a free field, which is an operator-valued 
distribution (in the sense of footnote~\ref{OpValDist}) in $x$ and
$e$, and which we denote by    
\begin{equation} \label{eqFieldIntegral}
\field(x,e) \doteq  \int_0^\infty ds_1\cdots \int_0^\infty ds_n\,
\field^\pt\big(x+(s_1+\cdots s_n)e\big). 
\end{equation}
Its associated intertwiner function is, after smearing with
$h(e)$, given by $\IntFct(p,h) =  t^n(p,h) \,\IntFctPt(p)$, that
is\footnote{We always understand $p\cdot e +i \eps$ in the sense of
  distributions: First smear against $h(e)$ and then take
  the limit $\eps \to 0$. 
}   
\begin{equation}  \label{eqIntFctIntegral}
  \IntFct(p,e) =  \frac{i^n}{(p\cdot e + i\eps)^n} \;\IntFctPt(p).
\end{equation}
Now for $e=e'+ie''$ in the tube $\calT_+$, the imaginary part $e''$ is
in the forward light cone~\cite{BrosMos}, hence $p\cdot e''$ is strictly
positive. Thus, the intertwiner is the boundary value 
of an analytic function on the tube. Further, the inverse Cauchy inequality 
$|p\cdot e''| \geq m (e''\cdot e'')^{1/2}$ holds, hence this function
is of moderate growth near the real boundary. Therefore,
$\IntFct(p,e)$ is an intertwiner function in the sense of
Definition~\ref{DefIntFct}. By Eq.~\eqref{eqvvcSt}, it is
self-conjugate if $\IntFctPt$ is. 
We have shown 
\begin{Prop}    \label{FieldIntegral}
The $n$-fold integral $\int_0^\infty ds_1\cdots \int_0^\infty
ds_n\field^\pt\big(x+(s_1+\cdots +s_n)e\big)$ exists as a
distribution in $(x,e)\in \RR^4\times \Spd$. It is the string-localized 
covariant free quantum field whose intertwiner function is given
by Eq.~\eqref{eqIntFctIntegral}. It is hermitean if $\field^\pt(x)$ is. 
\end{Prop} 

For later reference, we exhibit the two-point function of these line
integrals. 
Let $\field_1^\pt$ and $\field_2^\pt$ be free point-local fields for the
same particle type, let $\MTwoPt^\pt(p)$ be the on-shell part of its 
two-point function,  which is a polynomial~\cite{Weinberg}. 
Let, for $i=1$ and $2$, $\field_i(x,e)$ be the string-localized field
constructed 
from $\field_i^\pt$ by an $n_i$-fold line integral as in
Eq.~\eqref{eqFieldIntegral}. 
Recalling Eq.~\eqref{eqIntFctIntegral}, the on-shell part of the
corresponding two-point function 
$\TwoPt^\st(x-x',e,e') \doteq 
\lsp \Omega, \field_1(x,e)\field_2(x',e')\Omega \rsp$ 
is then given by
\begin{align} \label{eq2PtPtSt}
\MTwoPt^\st(p,e,e') = \frac{i^{n_2-n_1}\, \MTwoPt^\pt(p)}{(p\cdot
  e-i\eps)^{n_1}(p\cdot e'+i\eps)^{n_2}}.  
\end{align}
\subsection{Scaling degrees and massless limits} \label{secScalDeg}
The scaling degrees of the two-point functions involved in a given model
are decisive for the renormalizability of the model. 
The scaling degree of a two-point function $w(x-x')$ quantifies its
singular behaviour at coinciding points $x=x'$: scaling degree $\omega$ means
basically that it scales like $\lambda^{-\omega}$ under $x\mapsto
\lambda x$.       
The essential point is that the degree of freedom one has in giving a rigorous 
meaning to certain distributions which appear in the perturbation
series (see Outlook) grows with the scaling degrees of the involved
two-point functions.\footnote{This is so for the following reason. The 
  distributions in question are, at $n^{\text{th}}$ order of
  Epstein-Glaser perturbation theory,
  time-ordered distributions $t_n(x_1-x_2,\ldots,x_{n-1}-x_n)$ which
  are fixed {\em a priori} only outside the set of coinciding
  arguments $x_1=\cdots =x_n$~\cite{EG}. Thus one may add to a given
  choice of $t_n$ partial derivatives of
  $\delta_n\doteq\delta(x_1-x_2,\ldots,x_{n-1}-x_n)$, but this must be
  done without increasing the scaling degree. Now a $k^{\text{ th}}$
  order derivative of $\delta_n$ has scaling degree $4(n-1)+k$, hence
  one may add derivatives only up to order $\omega -4(n-1)$, where
  $\omega$ is the scaling degree of $t_n$. But $\omega$ depends in an
  additive way on the scaling degrees of the involved two-point functions. 
} 
If these are too large, then the number of free parameters grows with
the order of perturbation, and the model is not renormalizable by
power counting.   

All the two-point functions considered in the present article are sums
of terms of the form~\eqref{eq2PtPtSt}, with $n_1, n_2$ and the degree
$d$ of the polynomial $\MTwoPt$ varying 
from term to term, but the sum $n_1+n_2-d$ 
invariant. Thus, each term is homogeneous in $p$ of degree
$d-n_1-n_2$: one expects naively that the scaling degree should be
$d-n_1-n_2+2$ after smearing in $e,e'$. (That is, each factor 
$1/(p\cdot e\pm i\eps)$ should lower the scaling degree by one.)  
We show in Prop.~\ref{ScalDeg} that this is indeed the case, without
the singularity in $p\cdot e=0$ destroying this behaviour. 
 %
The two-point functions of our string-localized potentials turn out to
have $d-n_1-n_2=0$, independent of the spin ({\em
  c.f.\ }Eqs.~\eqref{eqMMVB} and \eqref{eq2PtFctAF} below), hence they
have scaling degree $2$, as promised. 

Moreover, we show in Prop.~\ref{Massless} that the massless limit of a
two-point function of 
the form~\eqref{eq2PtPtSt} exists if $d-n_1-n_2\geq -1$. This holds
for our string-localized potentials, and of course for the field
strengths (which have $n_1=0=n_2$). However, it is noteworthy that for 
$s\geq 2$ the massless limits of the field strength and of our string-localized
potential apparently do not coincide with the massless field strength and the
massless string-localized potential of~\cite{PlaschkeYngvason},
respectively.  

On the other hand, our escort fields 
fall into the case of infrared divergence, $d-n_1-n_2\leq -2$ ({\em c.f.}
Eqs.~\eqref{eqMPhi21} and \eqref{eqMPhi20} below). In this case the limit 
two-point function $\TwoPt(\xi,e,e')$ is well-defined on functions
$f(\xi)$ whose Fourier transform vanishes of order $n_1+n_2-d-1$,
{\em c.f. }Prop.~\ref{Massless}. There exists of course an extension of  
$\TwoPt$ to {\em all} functions ({\em i.e.}, an extension of $\TwoPt$'s  
Fourier transform across momentum zero), and it is unique only up to
addition of terms of the form  
$c_{\mu_1\cdots \mu_k} \xi^{\mu_1}\cdots \xi^{\mu_k}$, where $0\leq k\leq
n_1+n_2-d-2$ (corresponding to the derivatives of a momentum space
delta function). But we expect that none of the extensions satisfies 
positivity, similar to the case of the two-dimensional free massless
scalar field. An interesting question is if, like in the
two-dimensional case~\cite{Abdalla}, one can nevertheless construct
well-defined exponentials of the fields, leading to continuous
superselection rules. For the escort field with spin one, we claim
that this is the case, {\em c.f.} Section~\ref{Massless1}.

We begin with explaining in detail our notion of ``scaling degree after smearing
in $e,e'$'', or ``scaling degree with respect to coinciding $x$-arguments''.
We first recall the point-local case. 
After identifying the two-point function as a distribution in the 
difference variable $\xi\doteq x-x'$, the set of coinciding arguments
corresponds to the origin in $\RR^4$. Let $u$ be a distribution on
$\RR^4$. The rescaled distribution $u_\lambda$, $0<\lambda\leq 1$, is
defined (in the usual informal notation) by 
$$ 
u_\lambda(\xi)\doteq u(\lambda \xi). 
$$ 
More formally, that means 
\begin{equation}  \label{equLambda} 
   \langle u_{\lambda}, f \rangle \doteq  \langle u, f^\lambda \rangle \quad 
\text{ with } \;f^\lambda(\xi) \doteq \lambda^{-4} \,
f(\lambda^{-1}\xi).  
\end{equation}
The {\em scaling degree of $u$} (with respect to the origin) is the
infimum of all those $\omega\in\RR$ for which
\begin{equation} \label{eqSd}
\lambda^\omega \, \langle u_{\lambda},f\rangle  
\underset{\lambda \to 0}{\longrightarrow} 0 \qquad \text{ for all }
f\in \calD(\RR^4) .   
\end{equation}
(If there is no such $\omega$, then the scaling degree is said to be
infinite.)    

We now consider our string-localized two-point functions, which  are
distributions $u(\xi,e,e')$ on $\RR^4\times \Spd\times
\Spd$. The distribution $u_\lambda$ which arises from rescaling of $u$
along $\RR^4$ is defined by  
$$
u_{\lambda}(\xi,e,e')\doteq u(\lambda \xi,e,e').
$$
By the {\em scaling degree of $u$ after smearing in the
$e$-variables} we mean the infimum of all those $\omega\in\RR$ for which   
\begin{equation} \label{eqSdSt}
\lambda^\omega \, \langle u_{\lambda},f\otimes h\otimes h'\rangle  
\underset{\lambda \to 0}{\longrightarrow} 0 \qquad \text{ for all }
f\in \calD(\RR^4), \; h, h'\in\calD(\Spd).   
\end{equation}
In the literature~\cite{BruFred00,SchulzPhD}, this would be called the
scaling degree of $u$ with respect to the sub-manifold 
$\{0\}\times \Spd \times \Spd$.  
The scaling degrees with respect to various other sub-manifolds relevant
to the definition of time-ordered products are analyzed
elsewhere~\cite{MundSantos}.    
\begin{Prop}[Scaling degree] \label{ScalDeg}
Let $\TwoPt(\xi,e,e')$ be a two-point function whose on-shell part is 
of the form~\eqref{eq2PtPtSt}, with $\MTwoPt^\pt$ a polynomial of degree $d$. 
Then the scaling degree of $\TwoPt$ after smearing in the
$e$-variables is the maximum of $0$ and $d+2-(n_1+n_2)$.
\end{Prop}
Of course, in the case $n_1=n_2=0$ this is just the well-known fact that the
point-local two-point function with on-shell part $\MTwoPt^\pt$ has
scaling degree $d+2$. 
\begin{Proof}
First assume that $\MTwoPt^\pt$ is {\em homogeneous} of degree $d$. 
Let $f$ be a test function on $\RR^4$, with inverse Fourier transform 
$\check{f}(p)$, {\em c.f.} Eq.~\eqref{eqFT}, and let
$h,h'\in\calD(\Spd)$. Observing that the inverse Fourier transform of 
$f^\lambda$ is just $\check{f^\lambda}(p)=\check{f}(\lambda p)$, that 
$d\mu_m(\lambda^{-1}p)=\lambda^{-2} d\mu_{\lambda m}(p)$ and that 
$t^n(\lambda^{-1}p,h) = \lambda^n t^n(p,h)$, one gets for the rescaled
two-point function $\TwoPt_\lambda$ 
\begin{align} 
\langle \TwoPt_\lambda, f\otimes h\otimes h'\rangle & = 
2\pi \int_{H_m^+} d\mu_m(p) \check{f}(\lambda p)  \MTwoPt^\pt(p)
\; \overline{t^{n_1}}(p,h)\, t^{n_2}(p,h')\label{eq2Ptfhh}\\
&= 
2\pi \lambda^{-d-2+n_1+n_2}\int_{H_{\lambda m}^+} d\mu_{\lambda m}(p) \check{f}(p)
\MTwoPt^\pt(p) \; \overline{t^{n_1}}(p,h)\,
t^{n_2}(p,h'). \label{eq2Ptfhh'} 
\end{align}
In a reference frame, $p_0=\omega_{\lambda m}(\bfp)$ and the measure
$d \mu_{\lambda m}(p)$ is given by Eq.~\eqref{eqdmum}. Then the integral
in~\eqref{eq2Ptfhh'} reads 
$\int d^3\bfp \, F_{\lambda}(\bfp)$, with 
\begin{equation}  \label{eqFlambda}
F_\lambda(\bfp) \doteq \frac{1}{p_0}\, \check{f}(p)
\MTwoPt^\pt(p) \; \overline{t^{n_1}}(p,h)\, t^{n_2}(p,h')
\Big|_{p_0=\omega_{\lambda m}(\bfp)} .
\end{equation}
Now the spatial components satisfy $|p_i|\leq \|\bfp\| \leq \omega_{\lambda
  m}(\bfp)$, hence $|p_\mu|\leq \omega_{\lambda m}(\bfp)$ and
$|\MTwoPt^\pt(p)|\leq c  \; \omega_{\lambda m}(\bfp)^d$. Using the
bound~\eqref{eqtphBd} and $|\check{f}(p)|
\leq c(1+\|\bfp\|^r)^{-1}$, we get the bound 
\begin{equation}  \label{eqwBd}
|F_\lambda(\bfp)| \, \leq \,
\varphi(\bfp) \; \omega_{\lambda m}(\bfp)^{d-N-1} \,\leq
\varphi(\bfp) \; \omega_{\lambda m}(\bfp)^{d-N+1} \, \|\bfp\|^{-2},  
\end{equation} 
where $N\doteq n_1+n_2$ and $\varphi$ is an $\lambda m$-independent smooth
function of fast decrease.  

Consider first the case $d-N+1\geq 0$. Then 
$\omega_{\lambda m}(\bfp) \leq \omega_{m}(\bfp)$ implies  
\begin{equation}  \label{eqwBd'}
|F_\lambda(\bfp)| \, \leq \,
\varphi(\bfp) \; \omega_{m}(\bfp)^{d-N+1} \, \|\bfp\|^{-2},  \qquad
\lambda \in [0,1], 
\end{equation} 
hence the family $F_\lambda$ has a dominating $\lambda$-independent
$L^2$ function. Thus, the integral in Eq.~\eqref{eq2Ptfhh'} is
uniformly bounded in $\lambda$. (We shall see in the proof of
Prop.~\ref{Massless} that it 
actually converges for $\lambda\to 0$.) 
From the factor in front of the integral in Eq.~\eqref{eq2Ptfhh'} one
then reads off that the scaling degree is $d+2-N$, as claimed.  

If $d-N +1 < 0$, the fact that $\omega_{m}(\kappa)\doteq 
(\kappa^2+ m^2)^{1/2}$ is larger than $m$ and $(\kappa+m)/\sqrt{2}$
implies the bound    
$$
\omega_{\lambda m}(\kappa)^{d-N+1}
\leq \sqrt{2} (\kappa+\lambda m)^{-1} (\lambda m)^{d-N+2}. 
$$
Substituting this into Eq.~\eqref{eqwBd} and then into Eq.~\eqref{eq2Ptfhh'} 
we find the bound  
\begin{equation} \label{eq2PtfhhBd}
|\langle \TwoPt_\lambda, f\otimes h\otimes h'\rangle | 
\, \leq\, 2 \pi \sqrt{2}\,m^{d-N+2} 
\,\int_0^\infty d\kappa \,\frac{\tilde{\varphi}(\kappa)}{\kappa+\lambda m},
\end{equation}
where $\tilde{\varphi}(\kappa)\doteq \int_{S^2}d\Omega(\theta,\phi)
\varphi(\kappa,\theta,\phi)$ with $(\kappa,\theta,\phi)$ the polar
coordinates of $\bfp$. (The $\lambda$ factors in Eq.~\eqref{eq2Ptfhh'} cancel.)
Divide the integration region into $(0,m)$ and $(m,\infty)$. The
latter part is uniformly bounded, while the first part is bounded by
$$
c\int_0^m d\kappa
(\kappa+\lambda m)^{-1}= c\,\ln\big(\frac{1+\lambda}{\lambda}\big)
\leq c \ln(2/\lambda). 
$$      
Thus, the integral has a bound of the form
$a+b\ln(\lambda^{-1})$. Since for any $\eps>0$, 
$\lambda^\eps \ln(\lambda^{-1})$ goes to zero for $\lambda\to 0$,
the bound~\eqref{eq2PtfhhBd} implies that the scaling degree is $0$.

If the on-shell two-point function $\MTwoPt^\pt$ is not homogeneous,
the lower degree monomials will contribute terms with lower 
scaling degrees by the same token. This completes the proof. 
\end{Proof}
We now discuss the massless limit of two-point functions of the
form~\eqref{eq2PtPtSt}. To this end, we need a preparatory Lemma.
\begin{Lem} \label{htilde}  
Let $h\in\calD(\Spd)$ and let $\tilde{h}$ be its Fourier transform as
defined in Eq.~\eqref{eqhtilde}. Then there is for every $N\in\NN$ a
constant $c=c_{h,N}$ such that for every $p=\big(\omega_m(\bfp),\bfp)$
in the mass shell $\Hyp$ there holds the bound
\begin{equation} \label{eqhtildeBd}
|\tilde{h}(p)|\leq \, c\, (1+\|\bfp\|)^{-N}. 
\end{equation}
\end{Lem}
(The upshot here is that the right hand side is independent of $m$.) 
\begin{Proof}
In the reference system with $p=(\omega_m(\bfp),\bfp)$, points in
$\Spd$ are of the form $e=(e^0,\rho(e^0)\bs{n})$, with
$e^0\in\RR$ and $\bs{n}$ in the unit sphere, and $\rho(x)\doteq
(1+x^2)^\half$. Then the measure $d\sigma(e)$ on $\Spd$ is given by 
formula~\eqref{eqMeasuredS}, and we have 
\begin{align*}
\tilde{h}(p) & = \int_{S^2}d\Omega(\bs{n}) \int de^0
\rho^2(e^0)\,h(e^0,\bs{n})\, e^{i\varphi(e,p)},& \varphi(e,p) &\doteq  
\omega_m(\bfp)e^0-\bfp\cdot \bs{n} \, \rho(e^0).
\end{align*}
The derivative of the phase function $\varphi$ with respect to $e^0$ is 
\begin{equation} \label{eqAlpha<} 
\partial_0 \varphi(e,p) = \omega_m(\bfp) -\bfp \cdot \bs{n}\,
\frac{e^0}{\rho(e^0)} \,\geq \, \|\bfp\|\, (1-\frac{e^0}{\rho(e^0)}),
\end{equation}
in particular strictly positive. 
Now $N$ partial integrations yield 
\begin{align*}
\tilde{h}(p) & = i^N \int_{S^2}d\Omega(\bs{n}) \int de^0
\, e^{i\varphi(e,p)}\, \Big\{\big(\frac{\partial}{\partial e^0}\circ
a(e,p)\big)^N\,  
  \rho^2(e^0)\,h(e^0,\bs{n})\Big\}; & a(e,p)&\doteq \big(\partial_0
    \varphi(e,p)\big)^{-1},
\end{align*}
where $\big(\frac{\partial}{\partial e^0}\circ a(e,p)\big)^N$ is 
the $N$-fold repetition of the differential operator
$\frac{\partial}{\partial e^0}\circ a(e,p)$.  
The factor in curly brackets is a sum of terms of the form 
$a_1\cdots a_N f$, where $a_i$ is an $e^0$-derivative $\partial_0^n a$
of $a(e,p)$ of some order $n\in\{0,\ldots,N\}$ and $f$ is an
$e^0$-derivative of the function $\varrho ^2 h$. 
If we can show that the partial $e^0$-derivatives of $a(e,p)$ satisfy
an $e$-independent bound
\begin{equation} \label{eqan<}
|\partial_0^n a(e,p)| \leq c \|\bfp\|^{-1} 
\end{equation}
for all $e$ in the support of $h$, then we are done, since
$h\in\calD(\Spd)$ and $\tilde{h}$ is a continuous function on $\RR^4$
(by standard arguments).  
It remains to prove \eqref{eqan<}. 
To this end we write $a=b^{-1}$ with $b(e,p)\doteq
\partial_0\varphi(e,p)$. 
Then $\partial_0^n a(e,p)$ is a sum of terms of 
the form $\frac{b_1\cdots b_k}{b^{k+1}}$, where $b_i$ is an
  $e^0$-derivative of $b(e,p)$ of some order in $\{1,\ldots,n\}$.  
Now for $k\geq 1$ there holds 
$$
\|\partial_0^k b(e,p)\| \leq \|\bfp\|\, \big|\partial_0^k
\frac{e^0}{\rho(e^0)}\big|  \, \leq \, c\, \|\bfp\|
$$
for all $e$ in the (compact) support of $h$. Further,
Eq.~\eqref{eqAlpha<} implies that for all $e$ in the support of
$h$ there holds 
\begin{equation} 
b(e,p) \,\geq \, c'\, \|\bfp\|. 
\end{equation}
These facts imply the bound ~\eqref{eqan<}, and the proof is complete. 
\end{Proof}
\begin{Prop}[Massless limit] \label{Massless}
Let $\TwoPt(\xi,e,e')$ be a distribution whose on-shell part has the
form~\eqref{eq2PtPtSt},  
with $\MTwoPt^\pt$ a homogeneous polynomial of degree $d$, with mass independent
coefficients. If $n_1+n_2-d\leq 1$, then its limit $m\to 0$ exists in
the sense of distributions.  
If $n_1+n_2- d> 1$, then its limit $m\to 0$ exists for all functions
$f\otimes h\otimes h'$ where the Fourier transform of $f(\xi)$
vanishes of order $n_1+n_2-d-1$. 
\end{Prop}
\begin{Proof}
We consider the two-point function $\TwoPt$ for mass $\lambda
m$,\footnote{That is to say, we consider the integral in 
  Eq.~\eqref{eq2PtGeneral} over the mass shell for mass $\lambda m$.
} with $m$ fix,
and consider the limit  $\lambda\to 0$. Its value on a function $f\otimes
h\otimes h'$ is given by $2\pi \int d^3\bfp F_{\lambda m}(\bfp)$, with
$F_{\lambda m}$ given by Eq.~\eqref{eqFlambda} in  the proof of
Prop.~\ref{ScalDeg}.   
We first show that $F_{\lambda m}(\bfp)$ converges point-wise 
for $\lambda \to 0$ if $\bfp\neq \bs{0}$. It suffices to show that
$t_{(\omega_{\lambda m}(\bfp),\bfp)}(h)$ converges to $t_{(\|\bfp\|,\bfp)}(h)$. 
Recall that 
$t_p(h)$ is given by the $n$-fold integral~\eqref{eqtnuhtilde} over
$\tilde{h}$. Now $\tilde{h}$ is
continuous by the standard argument, and by Lemma~\ref{htilde} there
is a $\lambda m$-independent dominating function 
$$
|\tilde{h}\big(s\omega_{\lambda m}(\bfp),s\bfp)\big)|
\leq c(1+s\|\bfp\|)^{-N} 
$$
for the $s_i$-integrations, $s = s_1+\cdots +s_n$. Therefore
$t_{(\omega_{\lambda m}(\bfp),\bfp)}(h)$ converges for $\lambda \to
0$, and the same holds for $F_{\lambda m}(\bfp)$.  
For  $d-n_1-n_2 + 1 \geq 0$, we have already established a dominating
function~\eqref{eqwBd'} for $F_{\lambda m}$ in the proof of
Prop.~\ref{ScalDeg}. Therefore the limit $\lambda\to 0$ of the
two-point function exists.  

Suppose now that $d-n_1-n_2 + 1 <0$ and that $\check{f}$ has a zero of
degree $D\doteq n_1+n_2 - d-1$. Then $\check f$ is of the form 
$$
\check{f}(p) = \sum_{\mu_1,\ldots,\mu_{D}=0}^3 p_{\mu_1}\cdots
p_{\mu_D}\, \varphi^{\mu_1\cdots \mu_D}(p). 
$$
But $|p_\mu|\leq \omega_{\lambda m}(\bfp)$ for $p$ on the $\lambda m$-mass
shell, hence 
$$ 
|\check f(p)|\leq \omega_m(\bfp)^{D}\, \Phi(p) \qquad \text{ for } \;
p=(\omega_{\lambda m}(\bfp),p)\in\Hyp,  
$$  
where $\Phi(p)\doteq \sum  |\varphi^{\mu_1\cdots \mu_D}(p)|$. Then in the
bound~\eqref{eqwBd} the function $\varphi(\bfp)$ is replaced by
$ \omega_{\lambda m}(\bfp)^{N-d-1}\, \psi(\bfp)$, where $\psi$ is a smooth
function of fast decrease. The $\omega_{\lambda m}$-factors cancel,
and we have a dominating function $|F_{\lambda m}(\bfp)|\leq
\psi(\bfp)\,\|\bfp\|^{-2}$. Thus the limit $\lambda \to 0$ exists, and the
proof is complete.  
\end{Proof}
\section{Spin one: Massive vector bosons}  \label{secSpinOne}
The vector field for free massive particles with spin $1$
which acts on the (Hilbert) Fock space over the corresponding single
particle space is the Proca field $\APt_\mu(x)$, named after its
inventor~\cite{Proca36}.  
Its two-point function, see 
Eq.~\eqref{eqMProca} below, has a term quadratic in the momenta which
is responsible for the bad high energy behaviour of the field
$\APt_\mu$: it has scaling dimension 2.    
The Proca field is divergence free, and hence satisfies the Proca
equation: 
\begin{equation} \label{eqProca}
\partial^\mu \APt_\mu(x)=0,\qquad   \partial_\mu F^{\mu\nu}(x) + m^2
        {\APt}^\nu(x)=0, 
\end{equation}
where $F$ is  the field strength, $F=d\APt$:   
\begin{equation} \label{eqdAptF}
F_{\mu\nu}(x) \doteq \partial_\mu \APt_\nu(x) -\partial_\nu \APt_\mu(x). 
\end{equation} 
The field strength has the same scaling dimension as the Proca field;
in fact, its on-shell two-point function is a homogeneous polynomial of degree
$4$.  

We construct a string-localized version $\ASt_\mu(x,e)$ of the spin-one
vector field acting in the same Hilbert space, which has scaling
dimension one after smearing in $e$, and which has
the same field strength as the Proca field, that is, it satisfies the
identity\footnote{The partial derivatives $\partial_\mu$ always refer
  to the $x$ variable.}  
\begin{equation} \label{eqdAF} 
\partial_\mu \ASt_\nu(x,e) -\partial_\nu \ASt_\mu(x,e) = F_{\mu\nu}(x). 
\end{equation}
We therefore call it the {\em string-local vector potential}. 
In fact, the two versions of the potential differ by the gradient
of a ``scalar'' quantum field $\phi(x,e)$, the so-called {\em escort
  field}: 
\begin{align} \label{eqBAdPhi} 
\ASt_\mu(x,e)=\APt_\mu(x)+\partial_\mu \phi(x,e). 
\end{align}
This string-localized field $\phi$ transforms as a scalar field, but
corresponds to spin-one particles.    

In the next subsection, we construct $\ASt_\mu(x,e)$ and $\phi(x,e)$ as line
integrals over the field strength and over the Proca field,
respectively.  
In Subsection~\ref{secIntOne} we calculate explicitly the
corresponding Wigner intertwiners, and prove the
relations~\eqref{eqdAF} and \eqref{eqBAdPhi}. Along the way, we also
verify the above well-known facts on the Proca field.   
In Subsection~\ref{BRST} we compare our construction with the gauge-
or BRST approach.  
We close the section with a comment on the massless limits. 
\subsection{Definition as line integrals over point-local
  fields}  \label{secLine}  
A formal solution to Eq.~\eqref{eqdAF}, in the spirit of the Poincar\'e
Lemma, is obtained by the line integral    
\begin{equation} \label{eqDefA} 
A_\mu(x,e) \doteq \int_0^\infty ds F_{\mu\nu}(x+se)e^\nu, 
\end{equation}
where $e\in\Spd$. In fact, inserting~\eqref{eqdAptF} into
$F_{\mu\nu}$ and using the formal identity  
\begin{equation} \label{eqFormalId}
\int_0^\infty ds \,e^\nu\partial_\nu \APt_\mu(x+se) = -\APt_\mu(x),
\end{equation}
one readily verifies that the two potentials $\ASt$ and $\APt$
formally differ by a gradient as anti\-ci\-pated in
Eq.~\eqref{eqBAdPhi}, where the field $\phi(x,e)$ is defined by    
\begin{equation} 
 \label{eqDefPhi}
\phi(x,e) \doteq \int_0^\infty ds \APt_\nu(x+se)e^\nu.
\end{equation}
Thus, $\ASt_\mu$ should indeed satisfy Eq.~\eqref{eqdAF}. 
Note that the integrals \eqref{eqDefA} and \eqref{eqDefPhi} exist by
Prop.~\ref{FieldIntegral}, and that the
identity~\eqref{eqFormalId} is rigorous because $\APt_\mu$ goes to
zero for large space-like arguments in the sense of matrix elements
between local states. 

Since the Proca field is divergence free and the escort field $\phi$
satisfies the Klein-Gordon equation (as any free field for mass $m$),
there holds the Gupta-Bleuler type relation 
\begin{equation}\label{eqdAPhi}
\partial^\mu \ASt_\mu(x,e) + m^2 \phi(x,e) = 0.
\end{equation}
An interesting further property of our vector potential is that it is
orthogonal to the string, 
\begin{equation}  \label{eqAeorthog}
\ASt_\mu(x,e) \, e^\mu = 0, 
\end{equation}
which follows from the anti-symmetry of $F_{\mu\nu}$.  
This is reminiscent of the axial gauge condition, with the difference however
that here it is automatically satisfied by construction, and that the field
must be considered a distribution in $e$.  
\subsection{Construction via Wigner intertwiners}  \label{secIntOne}
We calculate here explicitly the intertwiners for all fields
mentioned above, and verify the claimed relations between them, 
namely, Eqs.~\eqref{eqdAF}, \eqref{eqBAdPhi}, \eqref{eqdAPhi} and
\eqref{eqAeorthog}. 

To begin with, we shall use the following realization $\Done$ of the
spin one representation of $O(3)$. In a rest frame of $\bar p$, every $R\in
O(3)$ corresponds to a $4\times 4$ matrix of the form 
$$
R=
\left( 
\begin{matrix}
1 & 0 \\ 0^t & \underline{R}   \quad 
\end{matrix}
\right),
$$ 
where $\underline{R}$ is an orthogonal $3\times 3$ matrix.
We consider the representation $\Done$ as realized on the 
space $\mathfrak{h}^{(1)}\doteq \CC^3$ by the defining representation of
the rotations,  
$$
\Done(R) \doteq \underline{R}. 
$$  
The $\PT$ transformation is represented by the operator of component-wise
complex conjugation in $\CC^3$, 
\begin{equation} \label{eqD1PT}
\Done(-\unity) (z_1,z_2,z_3) \doteq (\bar z_1,\bar z_2,\bar z_3).  
\end{equation}
(The relations corresponding to \eqref{eqDsT} are satisfied since the rotations
commute with complex conjugation.) By irreducibility, this choice is
unique up to a factor. 
Then the $\PT$ transformation is represented on the one-particle space
$L^2(\Hyp;\CC^3)$ by  
\begin{equation} \label{eqUPT1} 
 \big(\Uone(-\unity)\psi\big) (p) =  \overline{\psi(p)}.
\end{equation}
The target space for vector fields is $\target\doteq \CC^4$ with basis
denoted by $\{e_{(0)},\ldots,e_{(3)}\}$, and $D(\Lambda)$ acts as the defining 
representation of $O(1,3)$, $D(\Lambda)z\doteq \Lambda z$.  

We first discuss the Proca field. Since it is
a point-local vector field, the corresponding Wigner 
intertwiner is $e$-independent. By what
has been said above, see Eq.~\eqref{eqIntInt} and the remark at the
end of Section~\ref{Int-Field}, it is given by $\IntFctAPt(p)\doteq
\hat\IntFct^\pt \circ 
B_p^{-1}$, where $\hat\IntFct^\pt$ is a linear map from $\CC^4$ to $\CC^3$
satisfying $ \underline{R}\circ \hat\IntFct^\pt = \hat\IntFct^\pt\circ R$ for all 
$R\in SO(3)$. If one wants the Proca field to be divergence free, then
the intertwiner must satisfy $\bar p^\mu\hat\IntFct^\pt_\mu=0$, i.e.,
$\hat\IntFct^\pt \bar p=0$. The restriction of $\hat\IntFct^\pt$ to $\bar p^\perp$
is then an intertwiner between two irreducible representations of
$SO(3)$ and therefore unique (up to a factor).  
It is given by 
$\hat\IntFct^\pt z \doteq i\, \underline{z}$,
where we have written $\underline{z}\doteq (z^1,z^2,z^3)$ if 
$z=(z^0,\ldots, z^3)$. The factor $i$ has been chosen so as to make
the intertwiner self-conjugate.\footnote{\label{Intrinsic}More
  intrinsically, $\target$  
  is the complexification of the space of Lorentz vectors, and
  $\Hlittle^{(1)}$ is the complexification of 
  the (Minkowski-) orthogonal complement of the reference momentum
  $\bar p$, with scalar product given by the negative of the
  sesquilinear extension of the Minkowski product, $\lsp c\otimes x,
  c'\otimes x'\rsp \doteq - \,\bar c c' x\cdot x'$. Then $\Done(R)$
  and $\D(R)$ both act as  
$c\otimes x\mapsto c\otimes Rx$ (where in the case of $\Done$, $x$ is
  restricted to be in $\bar p^\perp$), and the intertwiner $\hat\IntFct^\pt$
  is given by $\hat\IntFct^\pt (c\otimes x)\doteq i c\otimes Ex$, where $E$ is
  the (Minkowski-) orthogonal projector onto $\bar{p}^\perp$. The
  intertwiner property follows from the fact the $R$ commutes with the
  projection $E$.}   
As explained above, we now get a Wigner intertwiner by setting
$$
\IntFctAPt_\mu(p)\doteq \IntFctAPt(p) e_{(\mu)} \equiv \hat\IntFct^\pt  B_p^{-1} 
e_{(\mu)}. 
$$
We define $\APt_\mu(x)$ to  be the free field for this
intertwiner as in Eq.~\eqref{eqFreeField}, and verify the following
well-known facts:    
\begin{Lem} \label{Proca}
The Proca field $\APt_\mu(x)$ is hermitean, local, covariant and
parity covariant\footnote{{\em i.e.}, covariant under
the orthochronous Poincar\'e group.}. 
It is divergence-free, $\partial^\mu \APt_\mu =0$. 
It is the unique, up to unitary equivalence, free quantum field for
massive spin-one particles with these properties. Its on-shell
two-point function is given by   
\begin{equation}   
\MTwoPt_{\mu,\nu}^{\APt\APt}(p)=  -g_{\mu\nu} + 
\frac{p_\mu p_{\nu}}{m^2}. \label{eqMProca}
\end{equation}
\end{Lem}
\begin{Proof}
Covariance and locality follow from the properties of the Wigner intertwiner. 
The latter obviously satisfies the intertwiner relation also for the
parity transformation, hence the field is parity covariant.  
For the proof that $\APt_\mu$ is hermitean, we need to show that the
intertwiner function $\IntFctAPt(p)$ is self-conjugate.  
Since $D(-\unity)=-\unity$ and Lorentz transforms commute with complex
conjugation,  $\Lambda \bar z = \overline{\Lambda z}$, we have  
\begin{equation*}
{\IntFctAPt}^{c}(p)z = \Done(-\unity) \IntFctPt(p)D(-\unity)\bar z 
= - \overline{\IntFctAPt(p) \bar z} 
= i \, \overline{\underline{ B_p^{-1} \bar z}}
=  i \,\underline{ B_p^{-1} z} = \IntFctAPt(p) z,
\end{equation*}
which proves the claim. 
The fact that the Proca field is divergence-free follows from 
\begin{equation} \label{eqvp0}
\IntFctAPt_\mu(p)p^\mu = \IntFctAPt(p) p=\hat\IntFct^\pt B_p^{-1}p=\hat\IntFct^\pt \bar p=0.
\end{equation}
In order to calculate the two-point function, note that for
$x=(x^0,\underline{x})$ and $w=(w^0,\underline{w})$ in $\RR^4$ there  
holds
$$
\lsp \hat\IntFct^\pt x,\hat\IntFct^\pt w\rsp_{\CC^3} = \lsp
\underline{x},\underline{w} \rsp_{\CC^3}
= x^0w^0-x\cdot w,  
$$ 
where 
$x\cdot w$ is the Minkowski product.   
Using $(B_p^{-1} x)^0 = \bar p \cdot B_p^{-1} x/m= p\cdot x/m$, we get
\begin{equation}   \label{eqvpxvpw}
\lsp \IntFctAPt(p) x,\IntFctAPt(p) w\rsp_{\CC^3} = \frac{(p\cdot x) \,
  (p\cdot w)}{m^2} - x\cdot w.  
\end{equation}
Substituting into Eq.~\eqref{eqM}, this yields Eq.~\eqref{eqMProca}.  

Uniqueness follows from the remark at the end of
Section~\ref{Int-Field}, since the vector representation of the
Lorentz group is irreducible. 
\end{Proof}
The interwiner function $\IntFctF$ of the field strength $F_{\mu\nu}$
is read off from Eq.~\eqref{eqdAptF}: It is given by  
\begin{equation} \label{eqIntFctF}
\IntFctF_{\mu\nu}(p)= i\, (p_\mu \IntFctAPt_\nu(p) -p_\nu \IntFctAPt_\mu(p)).  
\end{equation}
The on-shell part of its two-point function $\lsp \Omega, F_{\mu\nu}(x)
F_{\alpha\beta}(x')\Omega \rsp$ comes out as 
\begin{equation} \label{eqMF} 
  M_{\mu\nu,\alpha\beta}^{FF}(p)= -p_\mu p_\alpha g_{\nu\beta} + p_\nu p_\alpha g_{\mu\beta}
  -  p_\nu p_\beta g_{\mu\alpha} + p_\mu p_\beta g_{\nu\alpha} .   
\end{equation}
(The terms $\sim p^4$ cancel). 

We now discuss the string-local vector potential $\ASt_\mu$ and the
escort field $\phi$. 
The escort field $\phi(x,e)$ is defined by the line
integral~\eqref{eqDefPhi}, and Prop.~\ref{FieldIntegral} states that its
Wigner intertwiner is given by  
\begin{equation} 
\IntFctPhi(p,e)  \doteq \frac{i(\IntFctAPt(p)e)}{p\cdot
  e+i\eps}\,.  \label{eqIntPhi}
\end{equation}
Similarly, the  vector potential $\ASt_\mu$ is defined by the line
integral~\eqref{eqDefA}, and by Prop.~\ref{FieldIntegral} its Wigner
intertwiner is given by   
\begin{equation} \label{eqIntAStF}
\IntFctA_\mu (p,e)\doteq i\, \frac{\IntFctF_{\mu\nu}(p)e^\nu}{p\cdot e
  +i\eps}. 
\end{equation}
Substituting $\IntFctF_{\mu\nu}(p)$ as in Eq.~\eqref{eqIntFctF}, this
yields 
\begin{equation} \label{eqIntFctMVB}
 \IntFctA_{\mu}(p,e) =   \IntFctAPt_\mu(p) - 
\frac{p_\mu}{p\cdot e+i\eps}\,\IntFctAPt(p)e. 
\end{equation}
Rewriting this as 
\begin{equation}  \label{eqIntFctMVB'}
\IntFctA_\mu(p,e) =  \IntFctAPt_{\mu} (p) \,+\,   ip_ \mu \; 
\IntFctPhi(p,e)  
\end{equation}
yields the relation~\eqref{eqBAdPhi}, namely, $\ASt_\mu=
\APt_\mu+\partial_\mu \phi$. 
Using Eqs.~\eqref{eqM} and \eqref{eqvpxvpw}, the on-shell two-point functions 
of $\ASt_\mu$ and $\phi$, as well as the mixed ones, come out as  
\begin{align}   
\MTwoPt_{\mu,\nu}^{\ASt\ASt}(p,e,e')&= -g_{\mu\nu}-
\frac{p_\mu p_{\nu}\; (e\cdot e')}{(p\cdot e-i\eps)  (p\cdot e'+i\eps)}\, 
+\frac{p_{\mu}e_{\nu} }{p\cdot e-i\eps} + \frac{p_{\nu}e'_\mu }{p\cdot
  e'+i\eps} \label{eqMMVB}  \\
\MTwoPt^{\phi\phi}(p,e,e') & = \frac{1}{m^2}- \frac{e\cdot e'}{(p\cdot e-i\eps)
    (p\cdot e'+i\eps)},  \label{eqMPhi}\\
M^{\ASt\APt}_{\mu,\mu'} (p,e)&=  -g_{\mu\mu'} + \frac{p_\mu
  e_{\mu'}}{p\cdot e-i\eps}  \label{eqMAB}  \\ 
M^{\ASt\phi}_{\mu}(p,e,e') &= -i \, \big( \frac{e'_{\mu}}{p\cdot
  e'+i\eps} - \frac{p_\mu \, e\cdot e'}{(p\cdot e-i\eps)(p\cdot e'+i\eps)}
 \big)  \label{eqMAPhi} \\
M^{\APt\phi}_{\mu} (p,e') &= i \, \big(\frac{p_{\mu}}{m^2} - 
\frac{e'_{\mu}}{p\cdot e'+i\eps}  \big) \label{eqMBPhi} 
\end{align}
Summarizing, we have 
\begin{Prop} \label{PropMVB}
The fields $A_\mu(x,e)$ and $\phi(x,e)$ are hermitean and
string-local~\eqref{eqFieldLoc}. They satisfy
covariance~\eqref{eqCovTens} with $D(\Lambda)=\Lambda$ and $\unity$, 
respectively, and parity covariance.    
Their 
two-point functions are given by Eq.s~\eqref{eqMMVB} and \eqref{eqMPhi}. 
They relate to the Proca field as in Eq.~\eqref{eqBAdPhi}. All three
fields are string-local relative to each other.  
Finally, $\ASt_\mu$ satisfies the ``axial gauge''
condition~\eqref{eqAeorthog} and the Gupta-Bleuler type
relation~\eqref{eqdAPhi}. 
\end{Prop}
By Prop.~\ref{ScalDeg}, the field $A_\mu$ has a better scaling
dimension than its point-like counterpart $\APt_\mu$, namely $1$.  
Note that the two-point function differs from that of the Krein
version (see next subsection) by the last three terms in
Eq.~\eqref{eqMMVB}. These terms restore positivity of the two-point function. 
It is also interesting to note that Eq.~\eqref{eqMPhi} shows that
$\phi$ is of the form found in~\cite[Prop.~4.3]{MSY2} with 
$F(e,p)=i(mp\cdot e)^{-1}$, cf.\ Eq.~(72) in~\cite{MSY2}. 
\begin{Proof}
(Relative) string-locality, covariance and hermiticity are consequences of
Prop.s~\ref{Locality} and \ref{FieldIntegral}. 
Relation~\eqref{eqdAF} of course follows from the
identity~\eqref{eqBAdPhi}, but can also be verified on the intertwiner
level from  the identities 
\begin{equation} \label{eqpvvF}
p_\mu \IntFctA_\nu (p,e)- p_\nu \IntFctA_\mu(p,e) \, = \, p_\mu
\IntFctAPt_\nu(p) - p_\nu \IntFctAPt_\mu(p) \, =\, -i \IntFctF_{\mu\nu}(p).  
\end{equation}
Relations \eqref{eqdAPhi} and \eqref{eqAeorthog} have been shown
already. 
\end{Proof}
\begin{Prop}[Uniqueness] \label{AUnique}
$\ASt_\mu$ is uniquely characterized by the following properties:
String-locality, covariance, relation~\eqref{eqdAF}, and the fact
that its two-point function 
has scaling degree two after smearing in the $e$-variables.
\end{Prop}
Note that in the massless case, the string-local vector potential is
already fixed by the first three properties, without the condition on
the UV behaviour~\cite{MSY2}.
\begin{Proof}
Let $\ASt_\mu(x,e)$ be a free field satisfying the stated properties,
and let $\IntFct(p,e)$ be its intertwiner.  
We wish to show that it coincides with the expression~\eqref{eqIntFctMVB}. 
To begin with, the relation $d\ASt=d\APt$ implies 
$$
p\wedge \big(v^{k}(p,e) - v^{pk}(p) \big) = 0,\quad k=1,2,3.
$$
Here we consider $v^k(p,e)\doteq v_\mu^{k}(p,e)e^{(\mu)}$ 
as a vector in $\CC^4$.  
Specializing to $p=\bar p\equiv (m,\bs{0})$ and noting that $\bar
p\wedge w=0$ implies $w=c \bar p$, the above identity implies that for
every $z\in\CC^4$ there holds   
\begin{equation}   \label{eqvvChiHat}
\hat \IntFct(e) z = \hat{\IntFct}^\pt z +  \bar p\cdot z\, \chi(e), 
\end{equation}
where $\hat \IntFct(e)\doteq \IntFct(\bar p,e)$ and
$\hat{\IntFct}^\pt\doteq \IntFctAPt(\bar p)$, and $\chi$ is a function
on $\Spd$ with values in $\CC^3$.  
The ``small intertwiner relations''~\eqref{eqIntInt} for
$\hat\IntFct(e)$ and for $\hat{\IntFct}^\pt$ imply that this function
must be invariant under rotations, 
\begin{equation} \label{eqChiInvar}
 \underline{R} \; \chi(R^{-1}e)= \chi(e) 
\end{equation}
for all $R\in SO(3)$ and $e\in\Spd$. Recalling the
relation~\eqref{eqIntInt'} between $\hat\IntFct(e)$ and
$\IntFct(p,e)$, Eq.~\eqref{eqvvChiHat} implies  
\begin{equation}  \label{eqvvChi}
\IntFct(p,e) z = \IntFctAPt(p) z +  p\cdot z\; \chi(B_p^{-1}e).  
\end{equation}
Conversely, if $\chi(e)$ satisfies the invariance
property~\eqref{eqChiInvar}, then $\IntFct(p,e)$ as defined above
satisfies the intertwiner relation.  
Summarizing, we have shown that $d\ASt=d\APt$, with $\ASt_\mu$
stringlocal, is equivalent to the relation~\eqref{eqvvChi}, with
$\chi$ satisfying~\eqref{eqChiInvar}. Locality of $\ASt_\mu$ is
equivalent to analyticity of $\chi$ in the tube $\calT_+$. Thus,
$\chi(p,e) \doteq \chi(B_p^{-1}e)$ is a Wigner intertwiner from the
trivial (scalar) representation to $\Done$. Such intertwiner is unique up
to multiplication with a distribution $F$, which is the boundary value
of a meromorphic function in the upper complex half
plane~\cite[Thm.~3.3]{MSY2}.    
Note that one such intertwiner, which is also self-conjugate, is given
by $\IntFctAPt(p)e$. Thus, the intertwiner for $\ASt_\mu$ must be of
the form  
$$ 
\IntFct_\mu(p,e)= \IntFctAPt_\mu (p)+ ip_\mu\,F(p\cdot e)\,\IntFctAPt(p)e.
$$
(We have extracted a factor $i$ for later convenience.) 
%
The on-shell part of the two-point function for this field comes out as 
\begin{multline*} 
M^{AA}_{\mu \mu'} = \frac{p_\mu p_\nu}{m^2}
\Big\{1-i\, \overline{F(p\cdot e)} p\cdot e +i F(p\cdot e') p\cdot e' +
\overline{F(p\cdot e)}F(p\cdot e') \big((p\cdot e)(p\cdot e') - m^2
e\cdot e'\big) \Big\} \\
+ ip_\mu e_\nu \overline{F(p\cdot e)} - ip_\nu e'_\mu F(p\cdot e') -g_{\mu\nu}.
\end{multline*}
Now enters the condition that the scaling degree of the two-point
function be two: It implies that the on-shell part must be bounded for
large $p$, and in particular that the expression in curly
brackets must fall off at least as $|p|^{-2}$ for large $p$. 
Now if $F$ is not just of the form $F(p\cdot e) = c \, (p\cdot e +
i0)^{-1}$, then all five terms in the expression in curly brackets are
linearly independent, and the first term (constant 1, which does not
fall off) cannot be killed by the 
other ones. Thus,  $F$ must be of the form $F(p\cdot e) = c \, (p\cdot e +
i0)^{-1}$, where $c$ is a complex coefficient. Then the expression in curly
brackets reads 
$$
1-i\bar c  + i c +|c|^2 -\frac{|c|^2 \, m^2 \, e\cdot e'}{(p\cdot
  e-i0)(p\cdot e'+i0)},   
$$
and the condition that this must fall off at least as $|p|^{-2}$ implies that  
$1-i\bar c  + i c +|c|^2 =0$. A further restriction comes from the
self-conjugate requirement. Note that the intertwiner functions
$\IntFctAPt(p)$ and $i p_\mu \IntFctAPt(p)e$ are
self-conjugate. This implies that the intertwiner $\IntFct(p,e)$ is 
self-conjugate if and only if $c=ir$, with $r\in \RR$. The quadratic
equation then reads $1-2r+r^2=0$, or $r=1$. Now $\IntFct(p,e)$ is fixed,
namely, given just by Eq.~\eqref{eqIntFctMVB}.   
\end{Proof}

\subsection{BRST invariance and string-independence} \label{BRST} 
The standard gauge theoretic approach to massive vector bosons has
been initiated by St\"uckelberg~\cite{Stueckelberg38} and
Pauli~\cite{Pauli41} in order to overcome the bad UV behaviour of the
Proca field. We recall its up-to date version as exposed in Scharf's
monograph~\cite{ScharfGauge}. 
One starts with a vector field ${\AK_\mu(x)}$, whose on-shell
two-point function is just the constant 
$-g_{\mu\nu}$.\footnote{One can also consider a one-parameter
  family of fields interpolating between $\AK_\mu$ and $\APt_\mu$,
  see~\cite{IZ}.} 
Obviously, this field has a good UV behaviour (scaling
dimension one), but acts in an in\-de\-finite metric (or $\underline{{\rm
    K}}$rein) space since its two-point function is not positive
definite. In order to divide out the unphysical states by a gauge 
principle, one first introduces even further unphysical degrees of
freedom: Namely, the so-called St\"uckelberg field $\phiK$, as well
as ghost- and anti-ghost fields $\uK$, $\tildeuK$. In the gauge chosen
in \cite{ScharfGauge}, they all have the mass of the Proca field. 
One introduces an infinitesimal gauge or BRST transformation $s$ 
satisfying   
\begin{equation}  \label{eqBRST}
s \AK_\mu = \partial_\mu \uK,\quad s\phiK = \uK, \quad s\uK=0, \quad
s\tildeuK =-(\partial\cdot \AK+m^2\phiK). 
\end{equation}
(We have normalized the fields in a way differing from the literature
in order to emphasize the analogy with our approach.) 
One verifies that the BRST transformation is nil-potent, $s\circ s=0$. The
observables are defined as the cohomology classes (kernel modulo
range) of $s$ and act on a Hilbert space~\cite{ScharfGauge}. Thus, all
unphysical degrees of freedom are divided out: This is the BRST  
version of the gauge principle.    
For example the operator $\partial^\mu \AK_\mu + m^2\phiK$ is in the image
of $s$, see the right equation in \eqref{eqBRST}, hence its class is
the trivial observable. This is a Gupta-Bleuler type condition. 
The Eqs.~\eqref{eqBRST} also imply that the operators 
$\partial_\mu\AK_\nu-\partial_\nu\AK_\mu$ and $\AK_\mu-\partial_\mu \phiK $ are  
in the kernel of $s$, in particular 
\begin{equation} \label{eqsAdphi=0}
s\big(\AK_\mu-\partial_\mu \phiK \big) = 0. 
\end{equation}
The corresponding classes are non-trivial observables; the class of
the former is the field strength $F_{\mu\nu}$, and the class of the
latter is nothing but the Proca field~\cite{Pauli41}.\footnote{In fact, the
  two-point function of the St\"uckelberg field $\phiK$ is that of a scalar
  field with ``wrong'' sign, {\em i.e.}, its on-shell part is
  $-1$. Using this, one readily verifies that the two-point function
  of $\AK_\mu-\partial_\mu \phiK$ really coincides with that of the
  Proca field.} 

We can formulate our string-local set up in complete analogy,
replacing gauge (or BRST) invariance of observables by
string-independence. The Krein operators $\AK_\mu$ and $\phiK$
correspond here to our string-local $\ASt_\mu$ and its escort field
$\phi$. The role of the BRST operator $s$ is here
taken over by the differential $\de$ on the manifold $\Spd$ of
space-like directions,  
\begin{equation} \label{eqde} 
(\de f)(x,e) \doteq \sum_{\alpha=0}^2 \frac{\partial}{\partial e^\alpha}f(x,e)\;
de^\alpha \, , 
\end{equation}
where $f$ is a function on $\RR^4\times \Spd$ and the functions $e^\alpha$ 
are local coordinates on $\Spd$. Then $e$-independence of a function
$f$ means just $\de f=0$, and our equation~\eqref{eqBAdPhi},
$\ASt_\mu-\partial_\mu \phi = \APt_\mu$, can be re-written in analogy
to Eq.~\eqref{eqsAdphi=0}, namely  
\begin{equation*} 
\de (\ASt_\mu-\partial_\mu \phi)=0. 
\end{equation*}
In order to further parallel the BRST transformation, we define a
string-localized field $u(x,e) \doteq \de\,\phi(x,e)$. Then one has, analogously 
to the equations~\eqref{eqBRST},   
\begin{equation*}
\de \ASt_\mu =  \partial_\mu u,\quad \de\phi = u,  \quad \de u\equiv
(\de)^2\phi=0.  
\end{equation*}
The Gupta-Bleuler type condition $\partial^\mu A_\mu+m^2 \phi
=0$ holds in our approach as an identity between operators, see
Eq.~\eqref{eqdAPhi}. 
Hence, the role of the anti-ghost $\tildeuK$ of the BRST setting can here be
taken over by any one-form $\tilde{u}=\sum_\alpha \tilde{u}_\alpha(x,e)
de^\alpha$ satisfying $\de \tilde{u}=0$. 

Summarizing, we replace the gauge principle, namely the
requirement that $s X = 0$ for an observable $X$, by string-independence of
local observables, namely  $\de X =0$. The same is required for the S-matrix.  
We have neither unphysical states nor
unphysical fields: All fields $\AK$, $\phiK$ and $\uK$ are replaced
here by physical fields in the Borchers class of the Proca field. By
``physical'', we mean that they 
act in a Hilbert space so that its correlation functions have a
probability interpretation.   

For later reference, we expose the on-shell restrictions of the
two-point functions of $u_\alpha u_\beta$ and $A_\mu u_\alpha$, where
$u_\alpha(x,e)\doteq \frac{\partial}{\partial e^\alpha}\phi(x,e)$:  
\begin{align}  
M^{u u}_{\alpha,\beta}(p,e,e') &= 
-\frac{g_{\alpha\beta}}{(p\cdot e)(p\cdot e')} 
+\frac{p_{\alpha}e_{\beta} }{(p\cdot e)^2(p\cdot e')} 
+\frac{p_{\beta}e'_\alpha }{(p\cdot e)(p\cdot e')^2} 
-\frac{p_\alpha p_{\beta}\; (e\cdot e')}{(p\cdot e)^2 (p\cdot
  e')^2}\, \label{eqMuu},   \\
M^{A u}_{\mu,\alpha}(p,e,e') &= i\Big(-\frac{g_{\mu\alpha}}{p\cdot e'}
+\frac{p_{\mu}e_{\alpha} }{(p\cdot e)(p\cdot e')} 
+\frac{p_{\alpha}e'_\mu }{(p\cdot e')^2} 
-\frac{p_\mu p_{\alpha}\; (e\cdot e')}{(p\cdot e) (p\cdot
  e')^2}\Big)\, \label{eqMAu}.   
\end{align}
(Here we have suppressed the $\pm i \eps$ prescriptions: Every term 
$p\cdot e$ has to be read as $p\cdot e-i\eps$, and similarly $p\cdot
e'\doteq p\cdot e'+i\eps$.
The above formulas can be found by applying partial $e$-derivatives 
to the two-point functions of $\phi\phi$ and $\ASt_\mu\phi$.) 
\subsection{Massless limits}   \label{Massless1}
According to Prop.~\ref{Massless}, our string-local vector
potential has a mass zero limit, namely, its two-point function
converges for $m\to 0$ to the one of the massless vector field
introduced in~\cite{MSY2}.  
On the other hand, the escort field $\phi$ does  not have a mass zero
limit, due to the $1/m^2$ term in its two-point
function~\eqref{eqMPhi}, just like the the Proca field. 
However, in the difference $\chi(x,e,e')\doteq \phi(x,e)-\phi(x,e')$
this term drops out, 
and the same holds for the differential $u_\alpha(x,e)$, see Eq.~\eqref{eqMuu}.  
Now notwithstanding the absence of any $1/m^2$ terms, there is still a problem
with the massless limit of the $uu$ two-point function: Its on-shell
part is homogeneous in $p$ of degree  $-2$, and hence by
Prop.~\ref{Massless} it can be smeared only with test 
functions whose Fourier transforms vanish at the origin of first order. 

Nevertheless the commutator of $u_\alpha$ with $\ASt_\mu(x,e)$ exists
in the massless limit, and so the adjoint action of
$e^{iu_\alpha(f,h)}$ on the algebra $\calA$ generated by the $\ASt_\mu$'s is 
well-defined in this limit. It is an automorphism of $\calA$ which is
localized, in the sense of Doplicher, Haag and Roberts, in the
space-like cone specified by the supports of $f(x)$ and $h(e)$.  
Of course if $\hat f(0)=0$ then this is, as a representation,
equivalent with the vacuum representation (the identity) since then
$u_\alpha(f,h)$ is a well-defined operator affiliated with $\calA$. But
we expect that it is inequivalent if $\hat f(0)\neq 0$. In fact, we
conjecture that this sector does not contain mass zero particles in
the sense of Wigner, but only infra-particles (see
\cite{S63,Bu86,FroMorStro}), in the sense that the unitary representation $U(x)$
of the translations contains, apart from the vacuum, only improper
eigenvectors of the mass operator with mass zero.    
Our conjecture is based on the observation that the measure 
$
d\mu_0(p) M^{uu}_{\alpha,\alpha}(p,h,h)$ 
is homogeneous in $\bfp$ of degree zero, just like the Lorentz invariant
measure on the zero mass shell in $1+1$ dimensions. Therefore the
expectation value of $U(x)$ in the state defined by $A\mapsto (\Omega,
e^{iu_\alpha(f,h)} A e^{-iu_\alpha(f,h)} \Omega)$ has (as a function
of $x$) the same IR-structure as the corresponding function of the
free scalar field in $1+1$ dimensions~\cite[Eq.~(4.25)]{Bu96}, where
it is known that its Fourier transform does not contain a singular
part concentrated on the mass zero shell~\cite{Bu96}. But this means
that there are no (zero mass) proper eigenstates of the mass operator. 

The same arguments apply to the escort field $\phi(f,h)$ if one restricts
to functions $h$ with total integral zero, $\int d\sigma(e) h(e)=0$. 
For then the $1/m^2$ term in the $\phi\phi$ two-point function vanishes, and the
other term has the same degree of homogeneity ($-2$) as the $uu$
two-point function.  
\section{Higher Spin}  \label{secSpinGen}
We consider now the case of arbitrary integer spin $s\geq 2$. 
As mentioned above, among the infinity of free point-local 
fields for spin $s$~\cite{Weinberg} acting in a Hilbert space there
are two tensor fields with 
optimal UV behaviour, namely, scaling dimension $s+1$: One of them is
a totally symmetric tensor $\PotPt_{\mu_1\cdots \mu_s}$ of rank $s$, which is 
trace- and divergence free, and transforms under the
Lorentz group according to the irreducible representation
$\D^{(s/2,s/2)}$~\cite{SinghHagen}. It is uniquely 
characterized by these properties up to unitary equivalence.  
Applying a linear differential operator of order $s$ to this field,
see Eq.~\eqref{eqdAFs}, one obtains the field strength tensor
$\F_{\mu_1\nu_1\cdots \mu_s\nu_s}$ of rank $2s$. 
It has surprisingly the same scaling dimension $s+1$, in
fact, its on-shell two-point function is a homogeneous polynomial of
degree $2s$. These facts are conveniently understood in our framework and
recovered below. 

We first recall the
representation for these particles and construct the intertwiners for
the point-local potential and the field strength tensor. 
Then we construct our string-localized potential mentioned in the
introduction, which is related to the field strength tensor by the same
relation~\eqref{eqdAFs}. It differs from the point-local potential by
derivative terms as stated in Eq.~\eqref{eqPotPotPts}, and has scaling
dimension one, as asserted in Theorem~\ref{MainThm}.  
\subsection{Single particle space} 
We take the representation $\Ds$ as realized 
on the space of symmetric trace-free three-tensors within the
$s$-fold tensor product $(\CC^3)^{\otimes s}$.  
On $(\CC^3)^{\otimes s}$ we consider the scalar product induced by
that of $\CC^3$, 
\begin{equation}\label{eqScalProdC3s}
(u_1\otimes\cdots \otimes  u_s, v_1\otimes \cdots \otimes v_s) \doteq 
(u_1,v_1)\cdots (u_s,v_s),
\end{equation}
using one and the same symbol. 
Let $g_3$ be the
metric tensor of the canonical scalar product in $\CC^3$ and let $\hat
g_3$  be its lift to a contra-variant tensor of rank two,  $\hat g_3 =
\sum_{i=1}^3 g_3^{ij} \, e_{(i)}\otimes e_{(j)}$ with respect to a basis
$\{e_{(1)},e_{(2)},e_{(3)} \}$ in $\CC^3$. 
The tensor $\hat g_3$ is characterized by
the fact that    
\begin{equation}    \label{eqKronecker}
\lsp \hat g_3 ,u\otimes v\rsp = \lsp \bar u, v\rsp 
\quad \text{ for all } u,v \in \CC^3,     
\end{equation}
and is obviously invariant under the tensor product representation
$R\otimes R$ of $O(3)$. 
A symmetric tensor $t\in(\CC^3)^{\otimes s}$ is trace-free iff 
\begin{equation} \label{eqTraceless3}
(g_3)_{ij} \, t^{iji_3\cdots i_s} = 0 
\end{equation}
for all $i_3,\ldots, i_s$. 
The orthogonal\footnote{
  Orthogonality (hermiticity) of this projector is not claimed in
  \cite{Grensing}, but can be easily verified. See
  also~\cite[Eq.~(C.1)]{GuttenbergSavvidy}, where the same projector
  is used and claimed to be hermitean.} projector $E$ from
$(\CC^3)^{\otimes s}$ onto the subspace of symmetric trace-free
three-tensors is given by~\cite{Grensing}  
\begin{multline}  \label{eqGrensing}
E \, u_1\otimes \cdots \otimes u_s = \frac{1}{s!} \sum_{\pi\in S_s}
\sum_{k=0}^{[s/2]} (-1)^k c_k \, (\overline{u_{\pi(1)}},u_{\pi(2)})\cdots
(\overline{u_{\pi(2k-1)}},u_{\pi(2k)}) \times \\ 
\times  \Eps u_{\pi(2k+1)}\otimes \cdots \otimes u_{\pi(s)}\, \hat
g_3^{\otimes k}.  
\end{multline}
Here, $\Eps$ is the projection onto the symmetric tensors,   
\begin{equation} \label{eqE+}
\Eps\; u_1\otimes \cdots \otimes u_s \doteq \frac{1}{s!}\sum_{\pi\in S_s}
u_{\pi(1)}\otimes \cdots \otimes u_{\pi(s)} ,
\end{equation}
where $S_s$ denotes the permutation group of $s$ elements, $[s/2]$ 
denotes the integer part of $s/2$ and the $c_k$ are certain specific
positive numbers calculated in~\cite{Grensing}, in particular $c_0=1$
e $c_1=\frac{1}{3}$.   
The space of symmetric trace-free tensors has dimension $2s+1$ and is
irreducible under the product representation of the rotation
group~\cite{Hamermesh}. We take it as our little Hilbert    
space $\Hlittle^{(s)}$: 
\begin{equation} \label{eqE}
\Hlittle^{(s)}\doteq E\; \big( \CC^3\big)^{\otimes s} . 
\end{equation}
The representation $\Ds$ of $O(3)$ is just the restriction of the
tensor product representation, 
\begin{equation} \label{eqDsR}
\Ds(R) \;E\,u_1\otimes\cdots \otimes u_s \doteq 
  E\,\Done(R) u_1\otimes\cdots \otimes \Done(R) u_s.
\end{equation}
An anti-unitary representer of the PT transformation $-\unity$ is given by:  
\begin{equation} \label{eqDsPT}
\Ds(-\unity) \; E\, u_1\otimes\cdots \otimes u_s \doteq
E\, \bar u_1\otimes\cdots \otimes \bar u_s  , 
\end{equation} 
where $\bar u_k$ means component-wise complex conjugation, see
Eq.~\eqref{eqD1PT}. 
Note that then $E$ is an intertwiner from the representation $\Done
\otimes \cdots \otimes \Done$ of the group generated by the rotations and the PT
transformation to $\Ds$, {\em i.e.}, Eq.~\eqref{eqDsR} holds for  
$R \in O(3)$ and for $R = -\unity$. 
This enables us to build up spin-$s$ Wigner intertwiners from spin-one 
Wigner intertwiners: 
\begin{Lem}  \label{Int-One-Many}
Let $v$ be a Wigner intertwiner from a representation $\D'$ of the
Lorentz group to $\D^{(1)}$, and let $E$ be an intertwiner from the
$s$-fold tensor product representation $\Done\otimes\cdots\otimes
\Done$ of the rotation group to $\Ds$.   
Then 
$$
v^{(s)}(p,e)\doteq E\,\circ \, v(p,e)\otimes\cdots \otimes v(p,e)
$$
is a Wigner intertwiner from the $s$-fold tensor product representation 
$\D'\otimes\cdots \otimes\D'$ of the Lorentz group to $\Ds$. If $E$
intertwines also the respective representers of the PT transformation
$-\unity$, then $v^{(s)}$ is self-conjugate if $v$ is.    
\end{Lem}
Instead of taking $s$ times the same intertwiner $v$, one may
obviously take different intertwiners. The proof of the lemma is
straightforward.     
We finally exhibit the scalar product in $\Hlittle^{(s)}$ for later
reference: 
\begin{multline}  \label{eqScalProdHs} 
\lsp E\, u_1\otimes \cdots \otimes u_s, E\, v_1\otimes \cdots\otimes
v_s \rsp  = \frac{1}{(s!)^2} \sum_{\substack{\pi,\sigma\in S_s}}\sum_{k=0}^{[s/2]} \,
\lsp u_{\pi(1)},\overline{u_{\pi(2)}}\rsp \cdots \,\lsp
u_{\pi(2k-1)},\overline{u_{\pi(2k)}}\rsp  \times \\
\times \lsp \overline{v_{\sigma(1)}}, v_{\sigma(2)}\rsp \cdots 
\lsp\overline{v_{\sigma(2k-1)}}, v_{\sigma(2k)}\rsp   
\, \lsp u_{\pi(2k+1)}, v_{\sigma(2k+1)}\rsp \cdots \lsp u_{\pi(s)}, v_{\sigma(s)}\rsp .
\end{multline}
\subsection{Point-local fields} 
The target space $\target$ for the tensor potential is 
the space of symmetric trace-free tensors in $(\CC^4)^{\otimes s}$, and
$\D(\Lambda)$ is the corresponding restriction of $\Lambda^{\otimes
  s}$, which is equivalent to
$\D^{(s/2,s/2)}$~\cite{GelfandMinlosShapiro}. For the field strength,   
$\target$ is an invariant subspace of $(\CC^4)^{\otimes 2s}$, and
$\D(\Lambda)$ is the corresponding restriction of $\Lambda^{\otimes 2s}$.
Let $\IntFctAPt_{\mu}(p)$ and $\IntFctF_{\mu\nu}(p)$ be the
intertwiners for the spin-one potential and field strength tensor,
respectively, and define  
\begin{align} \label{eqIntFctPts}
\IntFctsPt_{\mu_1\cdots\mu_s}(p) & \doteq E\, \IntFctAPt_{\mu_1}(p)\otimes
\cdots \otimes \IntFctAPt_{\mu_s}(p) \\
\IntFctsF_{\mu_1\nu_1\cdots\mu_s\nu_s}(p) & \doteq E\, \IntFctF_{\mu_1\nu_1}(p)\otimes
\cdots \otimes \IntFctF_{\mu_s\nu_s}(p) \label{eqIntFctFs},  
\end{align}
where $\IntFctAPt_{\mu}(p)$ and $\IntFctF_{\mu\nu}(p)$ are the
intertwiners for the spin-one potential and field strength tensor,
respectively. According to Lemma~\ref{Int-One-Many}, these are
self-conjugate Wigner intertwiners. 
Let $\PotPt_{\mu_1\cdots\mu_s}$ and $\F_{\mu_1\nu_1\cdots\mu_s\nu_s}$ be
the corresponding free   hermitean, local and covariant tensor fields. 
The field strength tensor $\F_{\mu_1\nu_1\cdots\mu_s\nu_s}$ obviously
has the (permutation) symmetry properties stated in the introduction. 
%
\begin{Lem}    \label{PotPt} 
The fields $\PotPt_{\mu_1\cdots\mu_s}$ and $\F_{\mu_1\nu_1\cdots\mu_s\nu_s}$ are
related as in Eq.~\eqref{eqdAFs}. The tensor potential
$\PotPt_{\mu_1\cdots\mu_s}$ is trace- and divergence-free,
\begin{equation}\label{eqTracelessDivfrees}
g^{\mu\nu}\PotPt_{\mu\nu\mu_3\cdots\mu_s}=0,  \qquad  
\partial^\mu\PotPt_{\mu\mu_2\cdots \mu_s}(x)=0.  
\end{equation} 
It is the unique, up to unitary equivalence,
free quantum field for massive spin-$s$ particles which is a
symmetric trace-free rank-$s$ tensor. 

Both fields have scaling dimension $s+1$. In fact, the on-shell
two-point function of $\F_{\mu_1\nu_1\cdots\mu_s\nu_s}$ is a homogeneous
polynomial of degree $2s$. 
\end{Lem}
\begin{Proof}
The fact that the potential $\PotPt$ has   
divergence zero follows as in the spin-$1$ case from the properties of the
spin-$1$ intertwiner function, see Eq.~\eqref{eqvp0}.
To prove tracelessness of $\PotPt$, note that $\IntFctAPt_\mu(\bar
p)=\IntFctPt e_{(\mu)}=ie_{(\mu)}$ if $\mu=1,2,3$ and $=0$ if
$\mu=0$. Hence 
\begin{equation} \label{eqTracelessInt2}
g^{\mu\nu} \IntFctAPt_\mu(\bar p)\otimes \IntFctAPt_\nu(\bar p)
=- g^{ij} e_{(i)}\otimes e_{(j)} = \hat g_3, 
\end{equation}
since $g^{ij}=-g_3^{ij}$. Hence the tensor $g^{\mu\nu}
\IntFctAPt_\mu(\bar p)\otimes \IntFctAPt_\nu(\bar p)\otimes
\IntFctAPt_{\mu_3}(\bar p)\otimes \cdots \IntFctAPt_{\mu_s}(\bar p)$
is orthogonal in $(\CC^3)^{\otimes s}$ to every trace-free tensor in
the sense of \eqref{eqTraceless3}, {\em i.e.} its projection 
$E$ onto the space of trace-free tensors is zero. This implies
tracelessness~\eqref{eqTracelessDivfrees} of $\APt_{\mu_1\cdots \mu_s}$.  

Uniqueness follows from the remark at the end of
Section~\ref{Int-Field}, since the representation of the Lorentz group
on symmetric trace-free tensors of rank $s$ is irreducible, namely,
equivalent to the representation
$\D^{(s/2,s/2)}$~\cite{GelfandMinlosShapiro}. 
To prove relation~\eqref{eqdAFs}, insert the explicit
form~\eqref{eqIntFctF} of $\IntFctF$ into the definition
\eqref{eqIntFctFs}. This yields  
\begin{align} 
&\IntFctsF_{\mu_1\nu_1\cdots\mu_s\nu_s}(p)  =  
i^s E\,\big(p_{\mu_1}\IntFctAPt_{\nu_1}(p)-p_{\nu_1}\IntFctAPt_{\mu_1}(p)\big)
\otimes \cdots \otimes
\big(p_{\mu_s}\IntFctAPt_{\nu_s}(p)-p_{\nu_s}\IntFctAPt_{\mu_s}(p)\big).  
 \label{eqIntFIntPts}     
\\   \nonumber
& \equiv i^s
\sum_{I\subset\{1,\ldots,s\}} (-1)^{|I|}
p_{\mu_{j_1}}\cdots p_{\mu_{j_l}}p_{\nu_{i_1}}\cdots p_{\nu_{i_k}} 
E\, \IntFctAPt_{\nu_{j_1}}(p)\otimes \cdots \otimes\IntFctAPt_{\nu_{j_l}}(p)
\otimes \IntFctAPt_{\mu_{i_1}}(p)\otimes\cdots\IntFctAPt_{\mu_{i_k}}(p),
\end{align}
where $k=|I|$,  $I=\{i_1,\ldots, i_k\}$, $I^c=\{j_1,\ldots,
j_l\}$, $l=s-k$. 
Since the projection of the tensor product of
$\IntFctAPt_\mu$-intertwiners is just the intertwiner for
$\PotPt_{\mu_{j_1}\cdots \nu_{i_k}}$, this proves Eq.~\eqref{eqdAFs}. 

The on-shell part of the two-point function 
of the field strength tensor is, by the general formula~\eqref{eqM}, given by 
$$ 
\lsp \IntFctsF_{\mu_1\nu_1\cdots\mu_s\nu_s}(p),
\IntFctsF_{\alpha_1\beta_1\cdots\alpha_s\beta_s}(p) \rsp\equiv 
\lsp \IntFctF_{\mu_1\nu_1}(p)\otimes
\cdots \otimes \IntFctF_{\mu_s\nu_s}(p), E\, \IntFctF_{\alpha_1\beta_1}(p)\otimes
\cdots \otimes \IntFctF_{\alpha_s\beta_s}(p)\rsp.
$$
Writing out this scalar product in $\HlittleS$ as in 
Eq.~\eqref{eqScalProdHs}, each term is a product of $s$ factors of the
form $\big( \IntFctF_{\mu_i\nu_i}(p), \IntFctF_{\alpha_j\beta_j}(p)\big)$ 
or $\big( \overline{\IntFctF_{\alpha_i\beta_i}(p)}, \IntFctF_{\alpha_j\beta_j}(p)\big)$ 
or the complex conjugate of the latter.  
Complex conjugation of the intertwiners only amounts to an overall
sign, and thus each factor is, up to a sign, just the on-shell
two-point function of 
the spin-one field strength tensor, which we already know to be a
homogeneous quadratic polynomial, see Eq.~\eqref{eqMF}. 
Thus, the on-shell two-point function of the spin-$s$ field strength tensor is
a homogeneous polynomial of degree $2s$, as claimed. 
In a similar way, one sees that the on-shell two-point function of the
tensor potential $\PotPt_{\mu_1\cdots\mu_s}$ is a (non-homogeneous)
polynomial of degree $2s$.  Prop.~\ref{ScalDeg} then implies that the 
scaling degree of the two-point functions is $2s+2$, {\em i.e.}, the scaling
dimension of the fields is $s+1$. 
\end{Proof} 
\subsection{String-localized fields} 
Our string-localized tensor potential $\Pot_{\mu_1\cdots \mu_s}(x,e)$ is
defined as the free field corresponding to the Wigner intertwiner   
\begin{equation} 
\IntFcts_{\mu_1\cdots\mu_s}(p,e) \doteq E\, \IntFctA_{\mu_1}(p,e)\otimes
\cdots \otimes \IntFctA_{\mu_s}(p,e) . \label{eqIntFcts}
\end{equation}
According to Lemma~\ref{Int-One-Many}, this is a (self-conjugate)
Wigner intertwiner from the $s$-fold tensor 
representation of the Lorentz group to $\D^{(s)}$. Thus,
$\Pot_{\mu_1\cdots \mu_s}$ is a string-localized, covariant and hermitean
quantum tensor field.   
Inserting the explicit formula~\eqref{eqIntFctMVB'}, one sees that the
Wigner intertwiner can be written as 
\begin{align} 
\IntFcts_{\mu_1\cdots\mu_s}(p,e)  &= 
\IntFctsPt_{\mu_1\cdots\mu_s}(p)+
 \sum_{j=1}^s i p_{\mu_j} \IntFctsEsc{s-1}_{\mu_1\cdots\hat \mu_j\cdots
  \mu_s}(p,e) 
+ 
i^2 \sum_{\substack{i,j\in\{1,\ldots,s\}\\i\neq j}} p_{\mu_i}p_{\mu_j}
\IntFctsEsc{s-2}_{\mu_1\cdots\hat \mu_i\cdots \hat \mu_j\cdots 
  \mu_s}(p,e)     \nonumber 
\\
& \qquad \cdots + i^s p_{\mu_1}\cdots p_{\mu_s} \IntFctsEsc{0}(p,e) ,  
\label{eqIntFctsDer}
\end{align}
where the hat means omitting of the corresponding index, with 
\begin{align*} 
\IntFctsEsc{k}_{\mu_1\cdots\mu_{k}}(p,e)  & \doteq 
\frac{i^{s-k}}{(p\cdot e+i\eps)^{s-k}}\, E\, 
\underbrace{\IntFctAPt(p)e\otimes \cdots \IntFctAPt(p)e}_{s-k\text{ times}}
\otimes \IntFctAPt_{\mu_1}(p)\otimes\cdots \otimes \IntFctAPt_{\mu_k}(p).  
\end{align*}
%
By Lemma~\ref{Int-One-Many} and Proposition~\ref{FieldIntegral}, this
is a (self-conjugate) Wigner intertwiner. We denote by
$\phi^{(s,k)}_{\mu_1\cdots \mu_k}$ the corresponding string-localized,
covariant and hermitean tensor field. Again by 
Proposition~\ref{FieldIntegral}, these ``escort fields'' can be 
written as the line integrals
\begin{align} 
\phi_{\mu_1\cdots \mu_k}^{(s,k)}(x,e)= \int_0^\infty dt_1\cdots \int_0^\infty dt_{s-k}\;
\PotPt_{\mu_1\cdots\mu_s}\big(x+(t_1+\cdots +t_{s-k})e\big) \,
e^{\mu_{k+1}}\cdots e^{\mu_s}.
\end{align}
The string-localized tensor potential $\Pot_{\mu_1\cdots \mu_s}$ and the
escort fields $\phi^{(s,k)}_{\mu_1\cdots \mu_k}$ satisfy the 
properties stated in Theorem~\ref{MainThm}: 
\begin{Prop} \label{Pots}
The string-localized tensor potential $\Pot_{\mu_1\cdots \mu_s}$ is related to
its point-local version and the escort fields
$\phi^{(s,k)}_{\mu_1\cdots \mu_k}$ as in Eq.~\eqref{eqPotPotPts}.  
Its two-point function has scaling degree two after smearing in the
$e$ variables, and has a massless limit. It is a potential for the field
strength $\F_{\mu_1\nu_1\cdots \mu_s\nu_s}$ in the sense of
Eq.~\eqref{eqdAFs}. Finally, it satisfies the ``axial gauge'' condition 
\begin{equation} \label{eqAxialGauges}
\Pot_{\mu\mu_2\cdots \mu_s}(x,e)\,e^\mu = 0.
\end{equation}
\end{Prop} 
(However, it is neither trace- nor divergence free.) Explicit
formulas for the two-point functions in the spin-two case are given in
Eqs.~\eqref{eqTwoPthh}, \eqref{eqMPhi21} and \eqref{eqMPhi20} below.   
\begin{Proof}
The relation~\eqref{eqPotPotPts} can be read off from
Eq.~\eqref{eqIntFctsDer}. 
Further, note that in Eq.~\eqref{eqIntFIntPts} one can replace  
each $\IntFctAPt_\mu(p)$ by $\IntFctA_\mu(p,e)$ due to the spin-one
relation~\eqref{eqpvvF}. 
This proves relation~\eqref{eqdAFs}, with $\PotPt$ substituted by
$\Pot$. 
To prove the statements on the two-point function, we consider its
on-shell part, namely   
$$
\lsp
\IntFcts_{\mu_1\cdots\mu_s}(p,e),\IntFcts_{\alpha_1\cdots\alpha_s}(p,e')
\rsp .
$$
As in the proof of Lemma~\ref{PotPt}, we find that it 
is given by a sum, each term of which is a product of $s$ factors of the
form $\lsp \IntFctA_{\mu_i}(p,e), \IntFctA_{\alpha_j}(p,e')\rsp$ 
or $\big( \overline{\IntFctA_{\alpha_i}(p,e')}, \IntFctA_{\alpha_j}(p,e')\big)$ 
or  $\big( \IntFctA_{\mu_i}(p,e), \overline{\IntFctA_{\mu_j}(p,e)}\big)$. 
The factors $(\IntFctA_\mu(p,e),\IntFctA_\alpha(p,e'))$ are just the on-shell
functions $\MTwoPt^{AA}_{\mu\alpha}(p,e,e')$ given in Eq.~\eqref{eqMMVB},
and the other factors are determined, using Eq.~\eqref{eqvpxvpw} and
$e\cdot e=-1$, as  
\begin{align}   \label{eqvvbar} 
\big( \overline{\IntFctA_\mu(p,e)},\IntFctA_\nu(p,e) \big) &=
-g_{\mu\nu} + \frac{p_\mu\cdot  e_\nu+p_\nu \cdot  e_\mu}{p\cdot e + i\varepsilon} + 
 \frac{p_\mu  p_\nu }{(p\cdot e + i \varepsilon)^2}  
\end{align}
and its complex conjugate. Note that all these factors are homogeneous
in $p$ of degree zero. Thus, each product of $s$ factors is also
homogeneous in $p$ of degree zero. More precisely, it is a sum of
terms of the form~\eqref{eq2PtPtSt}, with the degree of the 
polynomial $\MTwoPtpt(p)$ just equal to $n_1+n_2$ (the number of
factors $p\cdot e\pm i \eps$ or $p\cdot e'\pm i \eps$ in the denominator). 
Then Prop.~\ref{ScalDeg} implies that the scaling degree is two, and
Prop.~\ref{Massless} implies that the massless limit exists.
\end{Proof}

We finally wish to write our string-localized tensor potential as a
line integral over the field strength tensor, in analogy to the spin-one
case, see Eq.~\eqref{eqDefA}. 
\begin{Lem} \label{AFInt} 
There holds 
\begin{align} \label{eqPotIntegralF}
\Pot_{\mu_1\cdots\mu_s}(x,\spd)= 
\int_0^\infty dt_1\cdots \int_0^\infty dt_s\;
\F_{\mu_1\nu_1\cdots\mu_s\nu_s}\big(x+(t_1+\cdots +t_s)e\big) \,
e^{\nu_1}\cdots e^{\nu_s}.  
\end{align} 
\end{Lem}
\begin{Proof}
Inserting relation~\eqref{eqIntAStF} into the
definition~\eqref{eqIntFcts} of the Wigner intertwiner of
$\Pot_{\mu_1\cdots \mu_s}$ yields    
\begin{equation} \label{eqIntAF}
 \IntFcts_{\mu_1\cdots\mu_s}(p,e)=  \frac{i^s}{(p\cdot e+i\eps)^s}
E\, 
\IntFctF_{\mu_1\nu_1}(p)e^{\nu_1}\otimes\cdots\otimes\IntFctF_{\mu_s\nu_s}(p)e^{\nu_s} 
= \frac{i^s}{(p\cdot e+i\eps)^s}
\IntFctsF_{\mu_1\nu_1\cdots\mu_s\nu_s}(p)e^{\nu_1}\cdots e^{\nu_s}. 
\end{equation}
Using Eq.~\eqref{eqIntFctIntegral}, this implies the claim. 
\end{Proof} 
Note that the (equivalent) relations \eqref{eqPotIntegralF} and
\eqref{eqIntAF} imply that the $AA$ on-shell two-point function can be
written as   
\begin{align}  \label{eq2PtFctAF}
\MTwoPt^{AA}_{\mu_1\cdots\mu_s,\alpha_1\cdots\alpha_s}(p,e,e') & = 
\frac{\MTwoPt^{FF}_{\mu_1\nu_1\cdots\mu_s\nu_s,\alpha_1\beta_1\cdots\alpha_s\beta_s}(p)}{(p\cdot
  e-i\eps)^s(p\cdot e'+i\eps)^s} 
\; e^{\nu_1} \cdots e^{\nu_s} (e')^{\beta_1}\cdots (e')^{\beta_s}, 
\end{align} 
where $M^{FF}$ is the on-shell part of the $FF$-two-point
function. Since this is a homogeneous polynomial of degree $2s$, 
Props.~\ref{ScalDeg} and \ref{Massless} now confirm respectively our
statements that the $AA$-two-point function has scaling degree two, and that the
massless limit exists.  
%
\subsection{Explicit formulas for the spin-two case}
The spin-two particles may be interpreted as massive gravitons, and
the tensor potential $\hpt_{\mu\nu}$ ($\equiv \APt_{\mu\nu}$) could,
in the massless limit, model the quantum fluctuations of the metric field.
The field strength would then be (twice) the linearized Riemann
tensor $R_{\mu\nu\alpha\beta}$. It is interesting to note that the
Ricci tensor $R_{\mu\alpha}\doteq g^{\nu\beta}R_{\mu\nu\alpha\beta}$
coincides in this linearized context of free quantum fields with a
multiple of the potential:   
\begin{equation} 
R_{\mu\alpha} = \half\, m^2 \hpt_{\mu\alpha}, 
\end{equation}
due to the the ``harmonic gauge'' conditions
\eqref{eqTracelessDivfrees} and the Klein-Gordon equation
for $\hpt_{\mu\nu}$.

We give a list of the on-shell two-point functions for the point- and
string-local tensor potentials $\hpt_{\mu\nu}$ and $\h_{\mu\nu}$, as 
well as for the escort fields $\phi^{(2,1)}_\mu$ and $\phi^{(2,0)}$. All these
are calculated as in the proofs of Lemma~\ref{PotPt} and
Proposition~\ref{Pots}, using the fact that for $s=2$ the
projection~\eqref{eqGrensing} onto the symmetric trace-free tensors in
$(\CC^3)^{\otimes 2}$ is given by 
$$
E\, u\otimes v = \Eps u\otimes v - \frac{1}{3} (\bar u,v) \, \hat g_3.
$$
For the point-local potential, the on-shell two-point function is 
\begin{align} \label{eqTwoPtHpt}
M_{\mu\nu,\alpha\beta}^{\hpt\hpt}(p)  &=  \frac{2}{3} \frac{p_\mu p_\nu p_\alpha
  p_\beta}{m^4}   \; + \;   \frac{1}{2} \left( g_{\mu \alpha} g_{\nu \beta}+
g_{\nu\alpha}g_{\mu \beta} \right) - \frac{1}{3}
g_{\alpha\beta}g_{\mu\nu}  \\
& \;\; -\frac{1}{2}\left( \frac{p_\mu p_\alpha}{m^2} g_{\nu\beta} +
\frac{p_\nu p_\beta}{m^2}g_{\mu \alpha} + \frac{p_\mu
  p_\beta}{m^2}g_{\nu \alpha}  +  \frac{p_\nu p_\alpha}{m^2} g_{\mu \beta}\right) 
+ \frac{1}{3} \left( \frac{p_\mu p_\nu}{m^2}g_{\alpha\beta}
       + \frac{p_\alpha p_\beta}{m^2 }g_{\mu \nu} \right) . \nonumber
\end{align}
(This coincides with Eq.~(21) in~\cite{Veltman70}, if one takes
account of the different conventions by substituting their
$\delta_{\mu\nu}$ by $-g_{\mu\nu}$.)   
The on-shell part of the string-localized $hh$-two-point function is 
given by
\begin{align} \label{eqTwoPthh}
& M_{\mu\nu,\alpha\beta}^{\h\h}(p,e,e') = 
 \frac{p_\mu p_\nu p_\alpha p_\beta}{(p\cdot e )^2(p\cdot
   e')^2} 
\big((e\cdot e')^2 -\frac{1}{3}\big) \\ 
& \hspace{2ex} - \frac{p_\mu p_\nu
  p_\alpha}{(p\cdot e)^2(p\cdot
  e')}\big(e\cdot e' e_\beta 
+ \frac{1}{3}e_\beta'\big)  
- \frac{p_\mu p_\nu
  p_\beta}{(p\cdot e)^2(p\cdot
  e')}\big(e\cdot e' e_\alpha 
+ \frac{1}{3}e_\alpha'\big)  \nonumber \\ 
&\hspace{2ex} - \frac{p_\mu p_\alpha p_\beta}{(p\cdot
  e)(p\cdot e')^2}\big(e\cdot 
e' e_\nu' + 
\frac{1}{3}e_\nu\big)  
- \frac{ p_\nu p_\alpha p_\beta}{(p\cdot e)(p\cdot
  e')^2}\big(e\cdot 
e' e_\mu' + 
\frac{1}{3}e_\mu\big)  \nonumber \\ 
&\hspace{2ex} + \frac{p_\mu p_\nu}{(p\cdot e)^2} \big(e_\alpha
e_\beta+\frac{1}{3}g_{\alpha\beta}  \big)   
+ \frac{p_\alpha p_\beta}{(p\cdot e')^2} \big(e_\mu
e_\nu+\frac{1}{3}g_{\mu\nu}  \big) \nonumber\\ 
& \hspace{2ex} 
 + \frac{p_\mu p_\alpha}{(p\cdot e)(p\cdot e')} \big(\frac{1}{2}
(e_\beta e_\nu'+ e\cdot e' g_{\nu\beta}) -\frac{1}{3} e_\nu e_\beta'\big)
\; + \; \frac{p_\mu p_\beta}{(p\cdot e)(p\cdot
  e')} \big(\frac{1}{2} 
(e_\alpha e_\nu'+ e\cdot e' g_{\nu\alpha}) -\frac{1}{3} e_\nu
e_\alpha'\big) \nonumber\\ 
& \hspace{2ex} + \frac{p_\nu p_\alpha}{(p\cdot e)(p\cdot
  e')} \big(\frac{1}{2} 
(e_\beta e_\mu'+ e\cdot e' g_{\mu\beta}) -\frac{1}{3} e_\mu
e_\beta'\big)
\; + \; \frac{p_\nu p_\beta}{(p\cdot e)(p\cdot
  e')} \big(\frac{1}{2}  
(e_\alpha e_\mu'+ e\cdot e' g_{\mu\alpha}) -\frac{1}{3} e_\mu 
e_\alpha'\big) \nonumber\\ 
& \hspace{2ex}+ \frac{p_\mu}{p\cdot e}\big(\frac{1}{3}g_{\alpha\beta}e_\nu
-\frac{1}{2}(g_{\nu\alpha} e_\beta+g_{\nu\beta}e_\alpha) \big)  
+ \frac{p_\nu}{p\cdot e}\big(\frac{1}{3}g_{\alpha\beta}e_\mu
-\frac{1}{2}(g_{\mu\alpha} e_\beta+g_{\mu\beta}e_\alpha) \big) \nonumber\\ 
& \hspace{2ex}+ \frac{p_\alpha}{p\cdot e'}\big(\frac{1}{3}g_{\mu\nu}e_\beta'
-\frac{1}{2}(g_{\beta\mu} e_\nu'+g_{\beta\nu}e_\mu') \big)  
+ \frac{p_\beta}{p\cdot e'}\big(\frac{1}{3}g_{\mu\nu}e_\alpha'
-\frac{1}{2}(g_{\alpha\mu} e_\nu'+g_{\alpha\nu}e_\mu') \big) \nonumber\\ 
& \hspace{2ex}+ \frac{1}{2} \left( g_{\mu \alpha} g_{\nu \beta}+ g_{\nu
  \alpha}g_{\mu \beta} \right) - \frac{1}{3}  g_{\alpha \beta}g_{\mu
  \nu} . \nonumber
\end{align}
Here we have suppressed the $i\eps$ prescriptions: Every factor
$p\cdot e$ is understood as $p\cdot e-i\eps$, while $p\cdot e'\doteq
p\cdot e'+i\eps$.  
The on-shell parts $M^{\phi^{(2,1)}}_{\mu,\alpha}$ and
$M^{\phi^{(2,0)}}$ of the two-point functions for the fields $\phi^{(2,1)}_\mu$ and 
$\phi^{(2,0)}$ come out as follows (again suppressing the $i\eps$
prescriptions):    
\begin{align} \nonumber 
M^{\phi^{(2,1)}}_{\mu,\alpha}(p,e,e')& = \frac{p_\mu p_\alpha}{m^2}
\big(\frac{2}{3m^2} - \frac{e\cdot e'}{2 (p\cdot e) (p\cdot e')} \big) +  
\frac{p_\mu }{m^2} \big(\frac{e_\alpha'}{3p\cdot e'} -
\frac{e_\alpha}{2p\cdot e} \big)  
+ \frac{p_\alpha }{m^2} \big(\frac{e_\mu}{3p\cdot e} -
\frac{e_\mu'}{2p\cdot e'} \big)  \\ 
& \quad+\frac{1}{2} g_{\mu\alpha} \big(\frac{e\cdot e'}{(p\cdot e)(p\cdot e')}
-\frac{1}{m^2}\big) 
 +\frac{1}{2} \frac{e_\alpha e_\mu'}{(p\cdot e)(p\cdot e')}
-\frac{1}{3}\frac{e_\mu e_\alpha'}{(p\cdot e)(p\cdot e')}  \label{eqMPhi21} \\
M^{\phi^{(2,0)}}(p,e,e') & =  \frac{2}{3m^4}  -
\frac{1}{3m^2}\big(\frac{1}{(p\cdot e)^2}
+\frac{1}{(p\cdot e')^2}\big) - \frac{2(e\cdot e')}{m^2  (p\cdot
  e)(p\cdot e')}  \nonumber \\ 
& \quad +  \frac{1}{(p\cdot e)^2(p\cdot e')^2}\big((e\cdot
e')^2-\frac{1}{3}\big).  \label{eqMPhi20}
\end{align}
\section{Outlook on interacting models}\label{secFinal}
Our free string-localized potentials shall be used to construct
interacting models along the lines of Epstein and
Glaser~\cite{EG}. Let us recall this scheme in the point-local case.
One starts from a given set of particle types with corresponding free fields
$\field\in\{\APt_\mu,\Dirac,\ldots\}$, and an interaction Lagrangean
$\Lint$, which is a Wick polynomial of the free fields and describes the
coupling between the various particles.\footnote{The free part of the
  classical Lagrangean does not enter into the construction.}   
The aim is to construct the S-matrix $S$, as well as for each free field
$\field\in\{\APt_\mu,\Dirac,\ldots\}$ an interacting version
$\field_\Lint$ which interpolates between the incoming free field
$\field_{\text{in}}\equiv \field$ and the outgoing free field
$\field_{\text{out}}\equiv S\field S^{-1}$ in the sense of the LSZ
relations~\cite{Hepp65}.      
The perturbative solution in the Epstein-Glaser scheme uses Bogoliubov's
S-matrix, which assigns to a test function $g(x)$ and the given
interaction Lagrangean $\Lint$   
the formal series of operators 
\begin{equation} \label{eqBogSMat}
S(g \Lint):= 
\sum_{n=0}^\infty \frac{i^n}{n!}\int d x_1\cdots d x_n
\; g(x_1)\cdots g(x_n) \; T_n\Lint(x_1)\cdots \Lint(x_n) ,
\end{equation}
where $T_n \ldots $ denotes the time-ordered product. The time-ordered 
distributions are recursively fixed only outside the set of coinciding
arguments, and the extension into 
this set is unique only after specifying some normalization
constants (UV problem). This is done so as to satisfy physically
motivated (re-) normalization conditions.  
If the scaling dimension of $\Lint$ is larger 
than four, then this leaves an infinite number of free parameters in
the series~\eqref{eqBogSMat}, and the model is non-renormalizable.  
One gets the physical S-matrix if one considers the so-called
adiabatic limit where $g(x)$ goes to a constant (IR problem). This
limit exists if all particles are massive~\cite{EGAdiabatic}.
The interacting version $\field_{g\Lint}$ for a given free field $\field$
is constructed via Bogoliubov's formula: 
\begin{equation} \label{eqBogFormula} 
\field_{g\Lint}(f):=\frac{1}{i}\frac{d}{d\lambda}S(g\Lint)^{-1}
S(g \Lint +\lambda f \field)\, \big|_{\lambda=0}. 
\end{equation}
(Of course it is covariant only in the adiabatic limit $g\to $const.) 

We wish to use our string-localized potentials in this scheme, since they have
better scaling dimensions.  
A look at the proof of locality in \cite{EG} shows that the interacting field
$\field_{g\Lint}$ is string-localized if the free field $\field$ is
string-localized and the interaction Lagrangean $\Lint$ is
{\em point}-local. However,  if $\Lint$ is string-localized, then
$\field_{g\Lint}$ is generically completely de-localized, no matter if
$\field$ is point- or string-local. We claim that, based on the
relation~\eqref{eqPotPotPts}, we can nevertheless  construct models from a
string-localized interaction Lagrangean with the following desirable
properties: The $S$-matrix is  
string-independent, the interacting version of a point-local
observable is still point-localized, and the interacting version of
a charge-carrying field is string-localized. We believe further that
there are models of this type which are renormalizable, although their
point-like counterparts are non-renormalizable. 

We have proved these claims (apart from renormalizability) at lower
orders in the example of massive QED~\cite{Mund_MassiveQED}. 
As a first step, it is shown 
on the basis of an analysis of the singularity structure of  string-localized
two-point functions~\cite{Amancio}, 
that Wick products and time-ordered products of string-localized fields are
well-defined.  
The particle types in massive QED are the ``massive photon'', the
electron and the positron, and   
the coupling is described by the interaction Lagrangean
\begin{equation} \label{eqLQED} 
\LintPt(x)  \doteq  j^\mu(x) \APt_\mu(x), 
\end{equation}
where $j^\mu\doteq \; \Wickli \bar{\Dirac}\gamma^\mu \Dirac\Wickre$ is
the current operator and $\Dirac$ is the free Dirac field. 
Now the scaling dimension of $j^\mu$ is three and that of $\APt$ is two
(see Lemma~\ref{Proca}), hence that of $\LintPt$ is five. Thus the model is
non-renormalizable as it stands. 
Our way out, analogous to the BRST approach~\cite{ScharfGauge}, is to
replace $\LintPt$ by its string-localized version
\begin{equation} \label{eqLQEDSt} 
\LintSt(x,e)  \doteq  j^\mu(x) \ASt_\mu(x,e),   
\end{equation}
which has a better scaling dimension, namely four, see the remark after
Prop.~\ref{PropMVB}. (It is under current investigation if this 
implies renormalizability~\cite{CardosoMund}.)  
By Eq.~\eqref{eqBAdPhi} and current conservation,
$\partial_\mu j^\mu=0$, the two interaction Lagrangeans differ by the
divergence of the string-localized vector field 
$V_\mu(x,e)\doteq j^\mu(x)\phi(x,e)$, where $\phi$ is the escort
field~\eqref{eqDefPhi}: 
\begin{equation} \label{eqLStLPtdV} 
\LintPt(x) = \LintSt(x,e) - \partial_\mu V^\mu(x,e).
\end{equation}
Then the product $\LintPt(x_1)\cdots \LintPt(x_n)$ also differs from
the $n$-fold product of $\LintSt$ by derivative terms containing the
$V_\mu$. Now the crucial question is whether this fact survives the
time-ordering, in other words: if the time ordering of the $T$-products
$T_n\LintSt\cdots\LintSt V^\mu\cdots V^\nu$ 
can be defined so that 
``the derivatives can be taken out of the $T$-products'': 
\begin{equation} \label{eqTLLVV}
T\LintPt_1\cdots \LintPt_n \stackrel{!}{=} T\LintSt_1\cdots \LintSt_n +
\sum_{\substack{I\subset\{1,\ldots,n\}\\I\neq\emptyset}}   
(-1)^{|I|}\, \partial_{\mu_1}\cdots\partial_{\mu_k} TV_{i_1}^{\mu_1}\cdots
V_{i_k}^{\mu_k}\LintSt_{j_1}\cdots \LintSt_{j_{n-k}} . 
\end{equation} 
(Here we have written $I=\{i_1,\ldots,i_{k}\}$,
$I^c=\{j_1,\ldots,j_{n-k}\}$, $\partial_{\mu_i}= 
\frac{\partial}{\partial x_i^{\mu_i}}$, and
$W_i=W(x_i,e_i)$ for $W=\LintPt$, $\LintSt$ or $V^\mu$.\footnote{Each
  $\LintSt$ and $V^\mu$ has its own $e$, since the products on the
  r.h.s.\ of \eqref{eqTLLVV} cannot be readily restricted to
  coinciding $e$'s, as one sees from the two-point
  functions~\eqref{eqMMVB}, \eqref{eqMPhi} and \eqref{eqMAPhi}. The
  restriction does exist if one considers light-like $e'$s.}      
This is a (re-) normalization condition for the $T$-products
$T_n\LintSt\cdots\LintSt V^\mu\cdots V^\nu$, which we call  
{\em perturbative string-independence}. (The condition can be
formulated without mentioning the $\LintPt$, namely: the right hand
side of Eq.~\eqref{eqTLLVV} be independent of the $e$'s.) 
It is analogous to the condition of ``perturbative gauge invariance"
in~\cite{ScharfGauge}.   
If it can be satisfied for all $n$, then the Bogoliubov S-matrix
$S(g\LintSt)$ is independent of the $e$'s in the adiabatic limit,
$g\to$ const., since the boundary terms vanish. 
In~\cite{Mund_MassiveQED} it has been shown that perturbative
string-independence can be satisfied in lower orders in massive QED.  
 
Now due to the good UV behaviour of the string-localized $\LintSt$ and
the restrictiveness of \eqref{eqTLLVV}, we
dare the conjecture that the model is 
renormalizable under the condition of string-independence; more
precisely: If one requires perturbative string-independence (and other
renormalization conditions) at every order, then $S(g\LintSt)$ is unique after
specifying a finite number of parameters.   

If this is true, then one may 
summarize our strategy as follows: The apparently
string-dependent $S(g\LintSt)$ can be constructed so as to be in fact
independent of the strings via the conditions of perturbative
string-independence~\eqref{eqTLLVV}, and these select a finite number out of the
infinitely many possible ways to define the non-renormalizable
$S(g\LintPt)$.  

Similarly, it has been shown at lower orders~\cite{Mund_MassiveQED} that the
interacting versions of observable fields like the current and the
field strength, if constructed according to Bogoliubov's
formula~\eqref{eqBogFormula} with $\Lint=\LintSt$, are $e$-independent in the
adiabatic limit.   
On the other hand, this does not hold for the interacting Dirac
field $\psi_{\LintSt}$. Rather, the interacting version of $\psi$ constructed
with $\LintSt$ coincides with the interacting version of the formal
Wick power series $\!\Wickli\exp(i g \phi)\psi\!\Wickre$ constructed
with $\LintPt$, namely: 
We conjecture\footnote{This conjecture has been verified in lower
  orders~\cite{FelipeDiss,Mund_MassiveQED}.} that the time ordered
products can be defined such that, roughly speaking,    
\begin{equation} \label{eqDiracInt}
\psi_{g\LintSt} \simeq 
\big(\!\Wickli e^{ig\phi}\,\psi\Wickre \big)_{g\LintPt} 
\end{equation}
holds in the adiabatic limit $g\to$ const. This relation
is well-known in the context of gauge theory (see Eq.~(34)
in~\cite{Ruegg2004}). Here, it may be read as   
follows: The right hand side is string-localized (since the interaction
Lagrangean is point-local and the free Wick series $\Wickli\exp(i g \phi)\,
\psi\Wickre$ is string-local), but non-renormalizable by
power-counting due to the bad UV behaviour of $\LintPt$. On the other
hand, the left hand side is apparently  
completely de-localized (since the interaction Lagrangean is not 
point-local), but has the chance to be renormalizable due to its
better UV behaviour. If this is the case, then the (conjectured)
equivalence of the two sides of \eqref{eqDiracInt} shows that the
interacting Dirac field 
$\Dirac_{g\LintSt}$ is string-localized and renormalizable.   
We wish to emphasize again that such a straightforward construction
with $\LintK\doteq j^\mu \AK_\mu$, where $\AK_\mu$ is the Krein space
vector boson, leads to an unphysical interacting Dirac field
$\psi_{\LintK}$. (This is done for example in Scharf's
monograph~\cite{ScharfQED}.)  

We also conjecture that the massless limit of the corresponding
correlation functions exists, leading to a construction of a physical
charged Dirac field in QED. 
(Here, the $\phi(x,e)$ in
  $\Wickli\exp(i g \phi)\Wickre$  of Eq.~\eqref{eqDiracInt} must be
  integrated with a fixed function $h(e)$ whose total integral is
  zero, $\phi_h(x)\doteq \int d\sigma(e)h(e) \phi(x,e)$.\footnote{ Then on the
  r.h.s.\ of Eq.~\eqref{eqDiracInt} appears $\Wickli\exp(i g
  \phi_h(x))\psi(x)\Wickre$, which is not linear in $h$.})  
It is plausible that this string-local Dirac field
describes the electron as an infra-particle in the sense that the spectral
measure in its sector is absolutely continuous. This phenomenon is a well-known
consequence of Gauss' law~\cite{Bu86,FroMorStro}, and we expect it to
manifest itself in our setting for the following reason.   
The spectral measure is just the  
Fourier transform of its two-point function, which is of the form 
\begin{equation} \label{eqDirac2Pt}
\langle \psi_{g\LintSt}(x,h) \; \bar \psi_{g\LintSt}(x',h)  \rangle =
\langle \Wickli e^{ig\phi_h(x)}\Wickre\;\!\Wickli e^{-ig\phi_h(x')}\Wickre
\rangle \;  \langle   \psi(x) \bar \psi(x') \rangle \; + \; \ldots 
\end{equation}
plus higher order terms in the perturbation series, if one uses the
right hand side of \eqref{eqDiracInt} for the series. ($\phi$ and
$\psi$ are the free fields and $\langle \,\cdot \, \rangle$ is the vacuum
expectation value.) Thus, in this approximation the spectral measure
is the convolution of the spectral measures of $e^{ig\phi_h(x)}$ and of the Dirac
field. Now as indicated in Subsection~\ref{Massless1}, the
field $e^{ig\phi_h(x)}$ is expected to describe massless infra-particles, hence
its spectral measure should be absolutely continuous (apart
from the $\delta$ contribution from the vacuum state).    
But then the spectral measure of the  string-localized Dirac
field is the convolution of an absolutely continuous measure with
another measure, and therefore itself absolutely continuous~\cite{ReSiII}.  

In other models, the condition of perturbative string-independence is
quite restrictive: For example, in models with several species
(``colors'') of self-interacting massive vector bosons it seems to imply the
Lie-Algebra structure and to require the coupling to a Higgs field, in
a way that one necessarily arrives at the non-Abelian Higgs
model~\cite{Bert_PeculiaritiesMVB, Bert_BeyondGauge} --- however, in
contrast to the usual approach, from first principles: There is no
symmetry breaking, the Mexican hat potential is {\em not} put in by
hand but comes out, the only input being that the first order coupling
be tri-linear in the $\ASt$'s. The gauge principle is replaced by the
more fundamental principle of locality.  
Further, it seems that any coupling of the vector bosons to fermionic
fields must be chiral, {\em i.e.}, only left- (or only right-) handed  
fermions couple~\cite{GMV_Chiral}. These features are analogous to the results
of the BRST approach~\cite{ScharfGauge}.

However, the question if these models are equivalent when constructed in
the string- or in the BRST setting requires further
investigation, and the same holds for the question if the class of
models which can be constructed with our string-localized fields
differs from that of the BRST approach.    
\paragraph{Acknowledgements.}
JM is grateful to Bert Schroer for his pioneer's spirit which has been
stimulating the project from its beginning,  
and to Jos\'e M.~Gracia-Bond\'ia, Joseph C.~V\'arilly and Karl-Henning
Rehren for helpful discussions and valuable suggestions for the
manuscript. This research was  
generously supported by the program ``Research in Pairs'' of the
Mathematisches Forschungsinstitut at Oberwolfach in November 2015.  
JM also thanks the Fakult\"at f\"ur Physik of the
Georg-August-Universit\"at G\"ottingen for the warm hospitality in 2015.  
JM and EO have received financial support by the Brazilian research
agencies CNPq and FAPEMIG, respectively. We are also grateful to CAPES
and Finep. 
\appendix
\setcounter{equation}{0}
\setcounter{Thm}{0}
\renewcommand{\theequation}{\thesection.\arabic{equation}}
\renewcommand{\theThm}{\thesection\,\arabic{Thm}}
\section{A Poincar\'e-type lemma for symmetric tensor fields}
\label{secPoincare}  
The classical Poincar\'e Lemma states that, in a topologically trivial
region, every closed form is exact. We are interested here in a
similar statement for symmetric tensors. Consider the 
differential operator $P$ of order $s$ which associates to a symmetric tensor
$A_{\mu_1\cdots \mu_s}$ of rank $s$ in flat (Minkowski) space a tensor
$(PA)_{\mu_1\nu_1\cdots  \mu_s\nu_s}$ of rank $2s$ with the
(permutation) symmetry properties of the field strength tensor, 
\begin{equation} \label{eqDerivatives}
(PA)_{\mu_1\nu_1\cdots \mu_s\nu_s}(x) \doteq 
\sum_{I\subset\{1,\ldots,s\}}(-1)^{|I|}
\partial_{\mu_{j_1}}\cdots \partial_{\mu_{j_{|I^c|}}} 
\partial_{\nu_{i_1}}\cdots\partial_{\nu_{i_{|I|}}}
A_{\nu_{j_1}\cdots \nu_{j_{|I^c|}}\mu_{i_1}\cdots \mu_{i_{|I|}}}(x),   
\end{equation}
where we have written $I=\{i_1,\ldots,i_{|I|}\}$ and
$I^c=\{j_1,\ldots,j_{|I^c|}\}$ for the complement of $I$. For $s=1$,
this is just the exterior 
derivative, $(PA)_{\mu\nu}=\partial_\mu A_\nu-\partial_\nu A_\mu$, and
for $s=2$ it is the linearized relation~\eqref{eqRhPt'} between the
Riemann tensor and a perturbation to the metric, 
\begin{equation} \label{eqDerivative2}
(PA)_{\mu\nu\alpha\beta} = 
  \partial_{\mu} \partial_{\alpha} A_{\nu\beta} 
- \partial_{\nu} \partial_{\alpha} A_{\mu\beta} 
 - \partial_{\mu} \partial_{\beta} A_{\nu\alpha} 
 + \partial_{\nu} \partial_{\beta} A_{\mu\alpha}.
\end{equation}
We have 
\begin{Prop} \label{PoincareSym}
Let $A$ be a symmetric tensor field of rank $s$ which falls off rapidly. Then 
$PA=0$ if, and only if, there are tensor fields $\phi^{(k)}$ 
of rank $k$, $0\leq k\leq s-1$, such that $A$ is of the form 
\begin{align} \label{eqPotPotPts'}
A_{\mu_1\cdots \mu_s}(x)&= \sum_{\substack{I\subsetneq\{1,\ldots,s\}}} 
\partial_{\mu_{j_1}}\cdots \partial_{\mu_{j_{|I^c|}}}
\phi^{(|I|)}_{\mu_{i_1}\cdots \mu_{i_{|I|}}}(x).    
\end{align}
Here the sum goes over all proper subsets $I$ of $\{1,\ldots,s\}$,
and we have written $I=\{i_1,\ldots,i_{|I|}\}$ and
$I^c=\{j_1,\ldots,j_{|I^c|}\}$. The tensor fields $\phi^{(k)}$ may be
taken as\footnote{Recall our sum   convention, {\em i.e.\ }here the sum over
  $\mu_{k+1},\ldots, \mu_s$ is understood.}   
\begin{equation} \label{eqPhikAsCl}
\phi_{\mu_1\cdots \mu_k}^{(k)}(x)= -\int_0^\infty dt_1\cdots \int_0^\infty dt_{s-k}\;
A_{\mu_1\cdots\mu_s}\big(x+(t_1+\cdots +t_{s-k})e\big) \,
e^{\mu_{k+1}}\cdots e^{\mu_s}.
\end{equation}
\end{Prop}
\begin{Proof}
If $A$ is of the form~\eqref{eqPotPotPts'} with some set of tensor
fields $\phi^{(k)}$, then one verifies
readily that $PA$ is zero. Conversely, suppose that $PA$ is
zero. Define the fields $\phi^{(k)}$ by Eq.~\eqref{eqPhikAsCl}, with some
fixed vector $e$, and substitute them into the right hand side of
Eq.~\eqref{eqPotPotPts'}. Then the r.h.s.\ of
Eq.~\eqref{eqPotPotPts'} reads
\begin{equation*}
-\sum_{k=0}^{s-1} \int_0^\infty dt_1\cdots \int_0^\infty dt_{s-k}\;
\partial_{\mu_{j_1}}\cdots \partial_{\mu_{j_{s-k}}} A_{\mu_{i_1}\cdots
  \mu_{i_k}\nu_{j_1}\cdots \nu_{j_{s-k}}}\big(x+(t_1+\cdots
+t_{s-k})e\big) 
e^{\nu_{j_1}}\cdots e^{\nu_{j_{s-k}}}.
\end{equation*}
Now we use the identity
$$
f(x)=(-1)^k \int_0^\infty ds_1\cdots\int_0^\infty ds_k\,
\partial_{\alpha_1}\cdots \partial_{\alpha_k}
f\big(x+(s_1+\cdots +s_k)e\big)\, e^{\alpha_1}\cdots e^{\alpha_k}
$$ 
and get  
\begin{multline*}
-\sum_{k=0}^{s-1} (-1)^k \int_0^\infty dt_1\cdots \int_0^\infty dt_s\;
\partial_{\mu_{j_1}}\cdots \partial_{\mu_{j_{s-k}}}
\partial_{\nu_{i_1}}\cdots \partial_{\nu_{i_k}}A_{\mu_{i_1}\cdots
  \mu_{i_k}\nu_{j_1}\cdots \nu_{j_{s-k}}}\big(x+(t_1+\cdots +t_s)e\big) \times \\
\times e^{\nu_{i_1}}\cdots e^{\nu_{i_k}} \, e^{\nu_{j_1}}\cdots e^{\nu_{j_{s-k}}}. 
\end{multline*} 
Since $\{i_1,\ldots,i_k \}\cup \{j_1,\ldots,j_{s-k} \}\equiv  I\cup I^c =
\{1,\ldots,s\}$, the product  of the $e$'s is just $e^{\nu_1}\cdots
e^{\nu_s}$, independent of $k$, and we get 
\small 
\begin{multline*}
-\int_0^\infty dt_1\cdots \int_0^\infty dt_s 
\Big\{\sum_{k=0}^{s-1} (-1)^k \;
\partial_{\mu_{j_1}}\cdots
\partial_{\mu_{j_{s-k}}}\partial_{\nu_{i_1}}\cdots
\partial_{\nu_{i_k}} A_{\mu_{i_1}\cdots 
  \mu_{i_k}\nu_{j_1}\cdots \nu_{j_{s-k}}}\big(x+(t_1+\cdots
+t_s)e\big)\Big\} \times 
\\ \times e^{\nu_1}\cdots e^{\nu_s} .   
\end{multline*} 
\normalsize
Now by hypothesis ($PA=0$) the expression in curly brackets is 
\begin{equation*} 
- (-1)^s \partial_{\nu_1}\cdots \partial_{\nu_s}
A_{\mu_1\cdots \mu_s}\big(x+(t_1+\cdots+t_s)e\big) .
\end{equation*}
Thus, the r.h.s.\ of Eq.~\eqref{eqPotPotPts'} is 
\begin{equation*}
 (-1)^s  \int_0^\infty dt_1\cdots \int_0^\infty dt_s 
\partial_{\nu_1}\cdots \partial_{\nu_s}
A_{\mu_1\cdots \mu_s}\big(x+(t_1+\cdots+t_s)e\big)
e^{\nu_1}\cdots e^{\nu_s} ,    
\end{equation*}
which is of course just $A_{\mu_1\cdots \mu_s}(x)$, as claimed. This
completes the proof.
\end{Proof}
\providecommand{\bysame}{\leavevmode\hbox to3em{\hrulefill}\thinspace}
\providecommand{\MR}{\relax\ifhmode\unskip\space\fi MR }
\providecommand{\MRhref}[2]{%
  \href{http://www.ams.org/mathscinet-getitem?mr=#1}{#2}
}
\providecommand{\href}[2]{#2}


\begin{thebibliography}{10}

\bibitem{Abdalla}
E.~Abdalla, M.C. Abdalla, and K.~Rothe, \emph{Non-perturbative methods in
  two-dimensional quantum field theory}, World Scientific, Singapore, 1991.

\bibitem{BrosMos}
J.~Bros and U.~Moschella, \emph{Two-point functions and quantum fields in de
  {S}itter universe}, Rev.\ Math.\ Phys.\ \textbf{8} (1996), 324.

\bibitem{BruFred00}
R.~Brunetti and K.~Fredenhagen, \emph{Microlocal analysis and interacting
  quantum field theories: Renormalization on physical backgrounds}, Commun.
  Math. Phys. \textbf{208} (2000), 623--661.

\bibitem{BGL}
R.~Brunetti, D.~Guido, and R.~Longo, \emph{Modular localization and {W}igner
  particles}, Rev.\ Math.\ Phys. \textbf{14} (2002), 759--786.

\bibitem{Bu86}
D.~Buchholz, \emph{Gauss' law and the infraparticle problem}, Phys.\ Lett.
  \textbf{B 147} (1986), 331--334.

\bibitem{Bu96}
\bysame, \emph{Quarks, gluons, colour: Facts or fiction?}, Nucl.\ Phys.\
  \textbf{B 469} (1996), 333.

\bibitem{BuF}
D.~Buchholz and K.~Fredenhagen, \emph{Locality and the structure of particle
  states}, Commun. Math. Phys. \textbf{84} (1982), 1--54.

\bibitem{CardosoMund}
L.~T. Cardoso and J.~Mund, \emph{work in progress}.

\bibitem{Dirac55}
P.~A.~M. Dirac, \emph{Gauge-invariant formulation of quantum electrodynamics},
  Can.~J.~Phys. \textbf{33} (1955), 650.

\bibitem{EG}
H.~Epstein and V.~Glaser, \emph{The role of locality in perturbation theory},
  Annales Poincar\'e Phys.\ Theor. \textbf{A 19} (1973), 211--295.

\bibitem{EGAdiabatic}
\bysame, \emph{Adiabatic limit in perturbation theory}, Renormalization theory
  (Erice) (G.~Velo and A.~S. Wightman, eds.), NATO Advanced Study Institute
  Series, vol.~23, 1976, pp.~193--254.

\bibitem{FroMorStro}
J.~Fr\"ohlich, G.~Morchio, and F.~Strocchi, \emph{Charged sectors and
  scattering states in quantum electrodynamics}, Ann.\ Phys. \textbf{119}
  (1979), 241--284.

\bibitem{GelfandMinlosShapiro}
I.M.\ Gel'fand, R.A.\ Minlos, and Z.Ya. Shapiro, \emph{Representations of the
  rotation and {L}orentz groups and their applications}, Pergamon Press,
  London, 1963.

\bibitem{GMV_Chiral}
J.~M.\ Gracia-Bond\'{i}a, J.~Mund, and J.~C. V\'arilly, \emph{The chirality
  theorem}, work in progress.

\bibitem{Grensing}
G.~Grensing, \emph{Symmetric and traceless tensors on {M}inkowski space}, Rep.
  Math. Phys. \textbf{14} (1978), 19--26.

\bibitem{GuttenbergSavvidy}
S.~Guttenberg and G.~Savvidy, \emph{Schwinger-{F}ronsdal theory of abelian
  tensor gauge fields}, SIGMA \textbf{4} (2008), 061.

\bibitem{H96}
R.~Haag, \emph{Local quantum physics}, second ed., Texts and Monographs in
  Physics, Springer, Berlin, Heidelberg, 1996.

\bibitem{Hamermesh}
M.~Hamermesh, \emph{Group theory and its application to physical problems},
  Adison-Wesley, 1962.

\bibitem{Hepp65}
K.~Hepp, \emph{On the connection between the {LSZ} and {W}ightman quantum field
  theory}, Commun. Math. Phys. \textbf{1} (1965), 95--111.

\bibitem{Hormander}
L.~H\"ormander, \emph{The analysis of linear partial differential operators
  {I}}, Springer, Berlin, 1983.

\bibitem{IZ}
C.\ Itzykson and J.-B. Zuber, \emph{Quantum field theory}, McGraw-Hill,
  Singapore, 1985.

\bibitem{Mandelstam}
S.~Mandelstam, \emph{Quantum electrodynamics without potentials}, Ann.\ Phys.
  \textbf{19} (1962), 1--24.

\bibitem{Strocchi03}
G.\ Morchio and F.~Strocchi, \emph{Charge density and electric charge in
  quantum electrodynamics}, J.\ Math.\ Phys. \textbf{44} (2003), 5569--5587.

\bibitem{Mund_MassiveQED}
J.~Mund, \emph{String-localized massive vector bosons without ghosts and
  indefinite metric: The example of massive {QED}}, in preparation.

\bibitem{M01a}
\bysame, \emph{The {B}isognano-{W}ichmann theorem for massive theories}, Ann.\
  H.\ Poinc. \textbf{2} (2001), 907--926.

\bibitem{Mu_JoSch}
\bysame, \emph{An algebraic {J}ost-{S}chroer theorem for massive theories},
  Commun.\ Math.\ Phys.\ \textbf{315} (2012), 445--464.

\bibitem{MundSantos}
J.~Mund and J.~A.~dos Santos, \emph{Singularity structure of the two-point
  functions of string--localized free quantum fields}, in preparation.

\bibitem{MSY}
J.~Mund, B.~Schroer, and J.~Yngvason, \emph{String--localized quantum fields
  from {W}igner representations}, Phys.\ Lett.\ B \textbf{596} (2004),
  156--162.

\bibitem{MSY2}
\bysame, \emph{String--localized quantum fields and modular localization},
  Commun.\ Math.\ Phys. \textbf{268} (2006), 621--672.

\bibitem{Nikolov16}
N.~M. Nikolov, \emph{Renormalization of massive {F}eynman amplitudes and
  homogeneity (bassed on a joint work with {R}aymond {S}tora)}, Nucl.\ Phys.
  \textbf{B} (2016), http://dx.doi.org/10.1016/j.nuclphysb.2016.07.002.

\bibitem{NST}
N.~M. Nikolov, R.~Stora, and I.~Todorov, \emph{Renormalization of massless
  {F}eynman amplitudes in configuration space}, Rev.\ Math.\ Phys. \textbf{26}
  (2014), 1430002.

\bibitem{ONeill}
B.~O'Neill, \emph{Semi--{R}iemannian geometry}, Academic Press, New York, 1983.

\bibitem{Pauli41}
W.~Pauli, \emph{Relativistic field theories of elementary particles}, Rev.\
  Mod.\ Phys. \textbf{13} (1941), 203--232.

\bibitem{FelipeDiss}
F.~M. Pedrosa, \emph{Campos qu\^anticos localizados tipo-string}, Ph.D. thesis,
  Universidade Federal de Juiz de Fora, Brazil, in preparation.

\bibitem{PlaschkeYngvason}
M.\ Plaschke and J.~Yngvason, \emph{Massless, string localized quantum fields
  for any helicity}, J.\ Math.\ Phys. \textbf{53} (2012), 042301.

\bibitem{Proca36}
A.~Proca, \emph{Sur la th\'eorie ondulatoire des \'electrons positifs et
  n\'egatifs}, J. de Phys. et le Radium \textbf{7} (1936), 347--353.

\bibitem{ReSiII}
M.~Reed and B.~Simon, \emph{Methods of modern mathematical physics {II}},
  Academic Press, New York, 1975.

\bibitem{Ruegg2004}
H.~Ruegg and M.~Ruiz-Altaba, \emph{The {S}tueckelberg field}, Int.\ J.\ Mod.\
  Phys.\ A \textbf{19} (2004), no.~20, 3265--3347.

\bibitem{Amancio}
J.~A.~dos Santos, \emph{An\'alise das singularidades da fun\c{c}\~ao de dois
  pontos do campo qu\^antico escalar localizado tipo-string}, Master's thesis,
  Department of Physics, Universidade Federal de Juiz de Fora, Brazil, 2010.

\bibitem{ScharfQED}
G.~Scharf, \emph{Finite quantum electrodynamics}, Springer, Berlin, 1989.

\bibitem{ScharfGauge}
\bysame, \emph{Quantum gauge theories}, Wiley, New York, 2001.

\bibitem{S63}
B.~Schroer, \emph{Infrateilchen in der {Q}uantenfeldtheorie}, Fortsch. Phys.
  \textbf{173} (1963), 1527.

\bibitem{Bert_PeculiaritiesMVB}
\bysame, \emph{Peculiarities of massive vector mesons and their zero mass
  limits}, Eur.~Phys.~J.~C. \textbf{75} (2015), 365.

\bibitem{Bert_BeyondGauge}
\bysame, \emph{Beyond gauge theory: Positivity and causal localization in the
  presence of vector mesons}, Eur.~Phys.~J.~C. \textbf{76} (2016), 378.

\bibitem{SchulzPhD}
R.~M. Schulz, \emph{Microlocal analysis of tempered distributions}, Phd thesis,
  Georg-August University School of Science (GAUSS), G\"ottingen, 2014.

\bibitem{SinghHagen}
L.~P.~S.\ Singh and C.~R. Hagen, \emph{Lagrangian formulation for arbitrary
  spin. {I}. {T}he boson case}, Phys.\ Rev. \textbf{D 9} (1974), 898--909.

\bibitem{St}
O.~Steinmann, \emph{A {Jost-Schroer} theorem for string fields}, Commun. Math.
  Phys. \textbf{87} (1982), 259--264.

\bibitem{Steinmann84}
\bysame, \emph{Perturbative {QED} in terms of gauge invariant fields}, Ann.
  Phys. \textbf{157} (1984), 232--254.

\bibitem{Steinmann}
\bysame, \emph{Perturbative {Q}uantum {E}lectrodynamics and axiomatic field
  theory}, Springer, 2000.

\bibitem{SW}
R.~F. Streater and A.S. Wightman, \emph{{PCT}, spin and statistics, and all
  that}, W. A. Benjamin Inc., New York, 1964.

\bibitem{Stueckelberg38}
E.~C.~G. St\"uckelberg, \emph{Die {W}echselwirkungskr\"afte in der
  {E}lektrodynamik und in der {F}eldtheorie der {K}ernkr\"afte {I}, {II},
  {III}}, Helv.\ Phys.\ Acta \textbf{11} (1938), 225--244,299--312,312--328.

\bibitem{Veltman70}
H.\ van Dam and M.~Veltman, \emph{Massive and mass-less {Y}ang-{M}ills and
  gravitational fields}, Nucl.~Phys. \textbf{B~22} (1970), 397--411.

\bibitem{VarillyGracia}
J.~C.\ V\'arilly and J.~M. Gracia-Bond\'{i}a, \emph{Stora's fine notion of
  divergent amplitudes}, Nucl.\ Phys. \textbf{B} (2016),
  http://dx.doi.org/10.1016/j.nuclphysb.2016.05.028.

\bibitem{Weinberg64}
S.~Weinberg, \emph{Feynman rules for any spin}, Phys.~Rev. \textbf{133} (1964),
  B1318--B1320.

\bibitem{Weinberg}
\bysame, \emph{The quantum theory of fields {I}}, Cambridge University Press,
  Cambridge, 1995.

\bibitem{Zwanziger76}
D.~Zwanziger, \emph{Physical states in quantum electrodynamics}, Phys. Rev.
  \textbf{D 14} (1976), 2570.

\end{thebibliography}
\end{document}